\documentclass[11pt]{article}
\usepackage{amssymb}
\usepackage{amsfonts}
\usepackage{amsmath}
\usepackage{mathtools}  
\usepackage{bm}
\usepackage{latexsym}
\usepackage{epsfig}
\usepackage{bbm}
\usepackage{enumerate}
\usepackage{float}
\usepackage{color}
\usepackage{gensymb}
\usepackage{siunitx}
\usepackage{color}
\usepackage{hyperref}
\usepackage[round,authoryear]{natbib}
\usepackage[T1]{fontenc}
\usepackage[utf8]{inputenc}
\usepackage{authblk}
\usepackage{appendix}
\usepackage{blindtext}

\usepackage[
        a4paper,
        left=2cm,
        right=2cm,
        top=2.5cm,
        bottom=3cm]{geometry}

\begin{document}

\title{Estimating changes in temperature extremes from millennial scale climate simulations using generalized extreme value (GEV) distributions}

\author[1]{Whitney K. Huang}
\author[2]{Michael L. Stein}
\author[3]{David J. McInerney}
\author[4]{Shanshan Sun}
\author[4]{Elisabeth J. Moyer}


\affil[1]{Department of Statistics, Purdue University}
\affil[2]{Department of Statistics, University of Chicago}
\affil[3]{School of Civil, Environmental and Mining Engineering, University of Adelaide}
\affil[4]{Department of the Geophysical Sciences, University of Chicago}

\date{ }

  \maketitle

\begin{abstract} 
Changes in extreme weather may produce some of the largest societal impacts of anthropogenic climate change. However, it is intrinsically difficult to estimate changes in extreme events from the short observational record. In this work we use millennial runs from the Community Climate System Model version 3 (CCSM3) in equilibrated pre-industrial and possible future (700 and 1400 ppm $\mathrm{CO}_2$) conditions to examine both how extremes change in this model and how well these changes can be estimated as a function of run length. We estimate changes to distributions of future temperature extremes (annual minima and annual maxima) in the contiguous United States by fitting generalized extreme value (GEV) distributions. Using 1000-year pre-industrial and future time series, we show that warm extremes largely change in accordance with mean shifts in the distribution of summertime temperatures. Cold extremes warm more than mean shifts in the distribution of wintertime temperatures, but changes in GEV location parameters are generally well-explained by the combination of mean shifts and reduced wintertime temperature variability. For cold extremes at inland locations, return levels at long recurrence invervals show additional effects related to changes in the spread and shape of GEV distributions. We then examine uncertainties that result from using shorter model runs. In theory, the GEV distribution can allow predicting infrequent events using time series shorter than the recurrence interval of those events. To investigate how well this approach works in practice, we estimate 20-, 50-, and 100-year extreme events using segments of varying lengths. We find that even using GEV distributions, time series of comparable or shorter length than the return period of interest can lead to very poor estimates. These results suggest caution when attempting to use short observational time series or model runs to infer infrequent extremes.
\end{abstract}

\section{Introduction} \label{sec1}
As the Earth's mean climate changes under increased concentrations of human-emitted greenhouse gases, the intensity and frequency of extreme weather conditions may change as well \citetext{\citealp{easterling2000}, chapter 11: \citealp{IPCC5} and references therein}. Extreme events, while rare by definition, can have large impacts on both human society and environmental systems: present-day weather damages are dominated by rare events that happen only every several decades or less \citep{NOAA2015}, and society's vulnerability to extreme events appears to be growing \citep{Kunkel1999}. In the United States, the frequency of climate and weather events with damages greater than \$1 billion appears to be increasing at around 5\% per year \citep{Smith2013}. It remains unclear to what extent long-term climate trends contribute to that rise, but these factors have led to extensive efforts to understand the relationship between long-term climate change driven by greenhouse gas forcing and potential changes in climate extremes.

Over the past two decades, numerous studies have sought to identify changes in temperature extremes both in observations \citep{easterling2000,shaby2012,parey2013,westra2013,lee2014,naveau2014} and in general circulation model (GCM) simulations of future climate  \citep{kharin2000,tebaldi2006,kharin2007,sterl2008,frias2012,kharin2013}. A number of studies find that extreme changes follow closely with changes in means and standard deviations \citep[e.g.][]{vries2012, parey2013}. \citet{parey2013}, for example, argue that changes in temperature extremes in station data from Eurasia and the United States are largely explained by changes in the mean and standard deviation of daily temperatures within the relevant season (summer for warm extremes and winter for cold extremes). Other studies may suggest more complex changes. \citet{barbosa2011} use quantile regression and clustering to show that the changes of the $5^{\text{th}}$ percentiles and the $95^{\text{th}}$ percentiles of daily air temperature over Central Europe are not the same as changes in medians, although they do not explicitly examine to what extent these differences can be explained by changing standard deviations. \cite{ballester2010} find evidence that accounting for changes in skewness in marginal distributions allows for a more accurate representation of changes in cold extremes than can be obtained from just considering changes in mean and standard deviation.

Analysis of changes in extremes is complicated by the fact that there is no unique definition of ``extremes''. One common definition is as exceedance of certain defined thresholds, with threshold values often defined based on past climate distributions \citep[e.g.][]{frich2002, alexander2006,tebaldi2006}. Analyses then evaluate changes in the frequencies at which these thresholds are surpassed. \cite{alexander2006}, for example, used percentile-based thresholds for various climate metrics from three decades of global gridded observations (1961-1990) and then considered changes in a subsequent period (1991-2003). They found a significant increase in the occurrence of annual warm nights (defined as the $90^{\text{th}}$ percentile of daily minimum temperature under the past climate) and a significant decrease in the occurrence of cold nights (defined as the $10^{\text{th}}$ percentile of daily minimum temperature under the past climate). However, those studies do not characterize changes in the shape of a distribution by making use of information on the magnitude of exceedances above (or below) the defined threshold.

\begin{figure}[H] 
\centering
\includegraphics[width=4in]{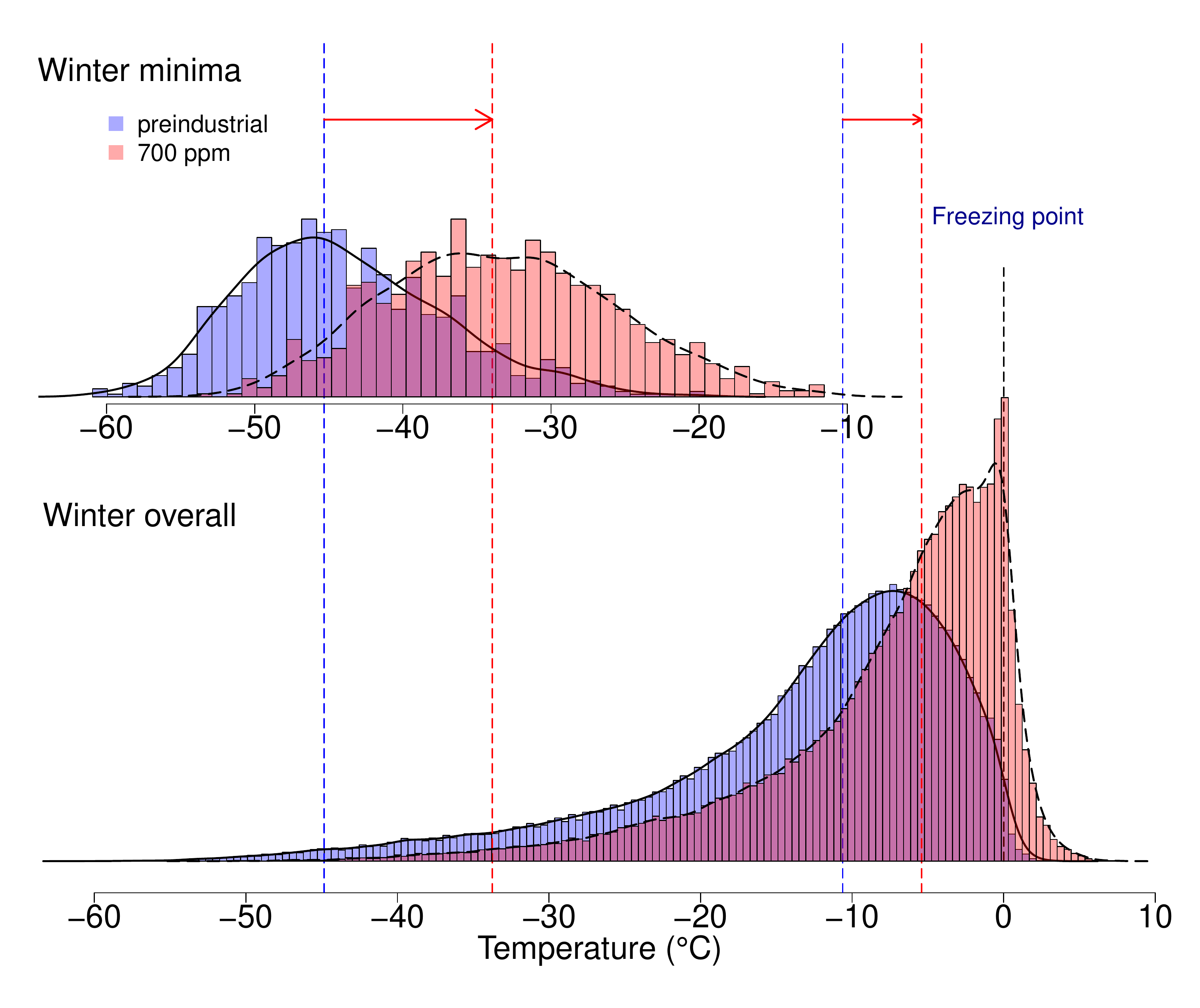}
\caption{Wintertime (December--February) temperatures from a location in Idaho from 1000-year CCSM3 model runs under pre-industrial and 700 ppm $\mathrm{CO}_2$ levels. \textbf{Lower}: marginal distributions of daily minimum temperatures in winter. \textbf{Upper}: distributions of winter seasonal minima. Note: 1) in both climate states, the distributions of seasonal minima are skewed to the right whereas their marginal distributions are skewed to the left; 2) in the future warmer climate state, variation in the marginal distribution decreases but variation in the seasonal minima increases; and 3) the median of the distribution of seasonal minima increases more than the median of the marginal distribution. We use seasonal instead of annual minima in this example so that the marginal and extreme distributions are based on the same data. The analogous plot for summertime maxima at this location is shown in Appendix Fig \ref{fig:ID_su}.}
\label{fig:ID_wi}
\end{figure}

An alternate and potentially more useful definition of ``extremes'' is based on occurrences in the far tail of the distribution of the quantity of interest. Extreme value theory (EVT) \citep{fisher1928,gumbel1958,coles2001,katz2002} provides a mathematical framework for studying these far tails. One common approach making use of EVT is based on ``block extremes", the maxima (minima) of some climate variable over given blocks of time. In climate studies, blocks of one year are common \citep[e.g.][]{zwiers1998}, although \cite{parey2013} use blocks of 45-50 days in order to get two blocks per season. Under some conditions, the magnitudes of extremes over sufficiently long blocks approximately follow a generalized extreme value (GEV) distribution \citetext{\citealp{gnedenko1943}, \citealp{leadbetter1983}, and see chapter 1 of \citealp{de2007} for details}. In these cases, the tail behavior of the distribution can be described by a functional form involving only three parameters: the location, scale, and shape parameters of the GEV distribution. For example, the widely used measure of extreme events, the $r$-year return level, can be expressed as a function of the GEV parameters (See Section~\ref{sec3} for further review). Annual extremes are directly relevant to important societal impacts; for example, the coldest temperature in a  winter can affect the abundance of the beetle population in the following spring \citep{shi2012,weed2013,zhang2013}. The behavior of annual extremes is difficult to discern from marginal distributions of daily temperatures. Figure~\ref{fig:ID_wi} illustrates how changes in block extremes can differ from changes in the underlying overall distribution, using seasonal temperature minima from millennial-scale climate model runs. Figure~\ref{fig:gev_changes} in Section~\ref{sec3.2} illustrates the effect on return levels of changing each individual GEV parameter.

The major limitation of the block extremes approach is that it discards all but the most extreme value within each block. In contrast, the peak over threshold (POT) method uses all data above a specified threshold. The POT method models exceedances over a (sufficiently high/low) threshold as a  generalized Pareto distribution (GPD) \citep{pickands1975,smith1989,davison1990}.
Point process theory as applied to exceedances over a threshold provides a unified way to derive
both GEV and GPD distributions \citep{resnick1987,smith1989}. While the POT method allows data on extremes to be used more efficiently, it requires choosing appropriate thresholds and declustering temporally dependent extremes. Threshold selection becomes complicated because temperature extremes show strong seasonality and temporal dependence. Furthermore, the temporal dependence in threshold exceedances means that it is not straightforward to convert inferences from the POT approach into inferences on annual extremes. We focus on the block extremes approach here because of its simplicity, its ease of interpretation, and its common usage in climate science.

An important advantage of using any EVT-based approach for the study of climate extremes is that it allows estimation of the probability of events that are more rare than the ``moderate extremes'' analyzed in threshold exceedance studies. The threshold exceedance study of \cite{alexander2006}, for example, used a threshold equivalent to the $90^{\text{th}}$ percentile of temperature calculated over a 5-day sliding window, which corresponds to events whose present mean recurrence interval is only 50 days. EVT allows examination of rare events of much longer recurrence intervals, and in principle even allows estimation of return levels longer than the observed range of the available time series together with a measure of statistical uncertainty. EVT has been widely applied in hydrology, for example, to estimate 100-year floods using several decades of data \citep{gumbel1958,katz2002}.

In the past two decades, EVT has been applied in numerous studies to climate variables, generally temperature and precipitation \citep[including][]{zwiers1998,kharin2005,kharin2007,sterl2008,frias2012,craigmile2013,kharin2013}. Studies have evaluated GEV distributions both in observational data and in output from general circulation models (GCMs) and regional climate models (RCMs). \cite{kharin2007,kharin2013} evaluated GCM ensembles from the CMIP3 and CMIP5 archives under several projected emission scenarios. In both studies, they found asymmetry in extreme changes, with high temperature extremes in most regions following changes in the mean summer temperature, while low temperature extremes warmed substantially more than mean winter temperatures.

Prior GEV studies are limited to some extent by two factors: the length and non-stationarity of the analyzed time series. For observational studies, existing data records are relatively short (often only several decades). Model runs may be longer, but the model runs used in these studies typically extend only around 100 years from the present. For some models and scenarios, ``initial condition ensembles'' are available, i.e. multiple runs of the same model and forcing scenario that differ only in their initial conditions.  Such ensembles obviously provide further information about extreme events, but they generally include only a few model realizations. The length of the time series has a large influence on the ability to detect changes in extremes, and a record that is too short can lead to large uncertainty in estimated return levels at long periods (see Section \ref{sec6}). Furthermore, at present and for the foreseeable future, the Earth's climate is not in equilibrium but is evolving (transient), so the GEV model must be extended to account for non-stationarity in climate extremes. The means of extension are not trivial. The most commonly used nonstationary GEV model \citep[e.g.][]{kharin2005} assumes that location and scale parameters change linearly in time and the shape parameter is time-invariant. A more flexible approach to modeling nonstationarity can be achieved by using generalized additive temporal structure of the parameters for extreme value distributions \citep{chavez2005, yee2007, heaton2011}. Given the limited information about extreme events, it is a challenging and important problem to decide on an appropriate degree of flexibility
in a nonstationary model.

In this study we avoid the limitations of short transient runs by using three long (millennial) climate model runs in which climate is fully equilibrated. Although numerical simulations of future climates provide only suggestions for possible changes in climate variables, not direct evidence of changes, they are important complements to observational studies. We use temperature output from the widely used Community Climate System 3 (CCSM3) model \citep{collins2006,yeager2006} at three different $\mathrm{CO}_2$ levels. For each scenario, the climate model is run for a sufficiently long warmup period to ensure the climate has fully responded to forcing changes, so that any time series of block extremes effectively forms a stationary sequence. Millennial stationary runs allow a more accurate determination of any changes in GEV distributions than do the shorter runs of the CMIP archives. The long model runs also allow us to assess, based on the max-stable property of GEV distributions (see Section \ref{sec5} for details), the appropriateness of the choice of annual blocks that have been traditionally used with shorter (century-scale) climate model runs. Finally, millennial runs allow us to evaluate empirically the sampling errors when estimating extreme return levels in shorter model runs, because we can use estimates based on entire long runs as ``ground truth''.

The paper is structured as follows: in Section~\ref{sec2}, we describe the climate model output used in this work; in Section~\ref{sec3}, we provide background for the univariate extreme value theory we employ; in Section~\ref{sec4}, we describe the changes in extreme value distributions and corresponding return levels, and compare them with changes in climate means. In Section~\ref{sec5}, we assess the assumption that annual blocks are sufficiently long for the GEV approximation to be valid, and in Section~\ref{sec6} we assess the sensitivity of estimates of return levels to the series length. We conclude with a discussion of the implications of these results.

\section{GCM output} \label{sec2}
\begin{figure}[H]
\centering
\includegraphics[width=6in]{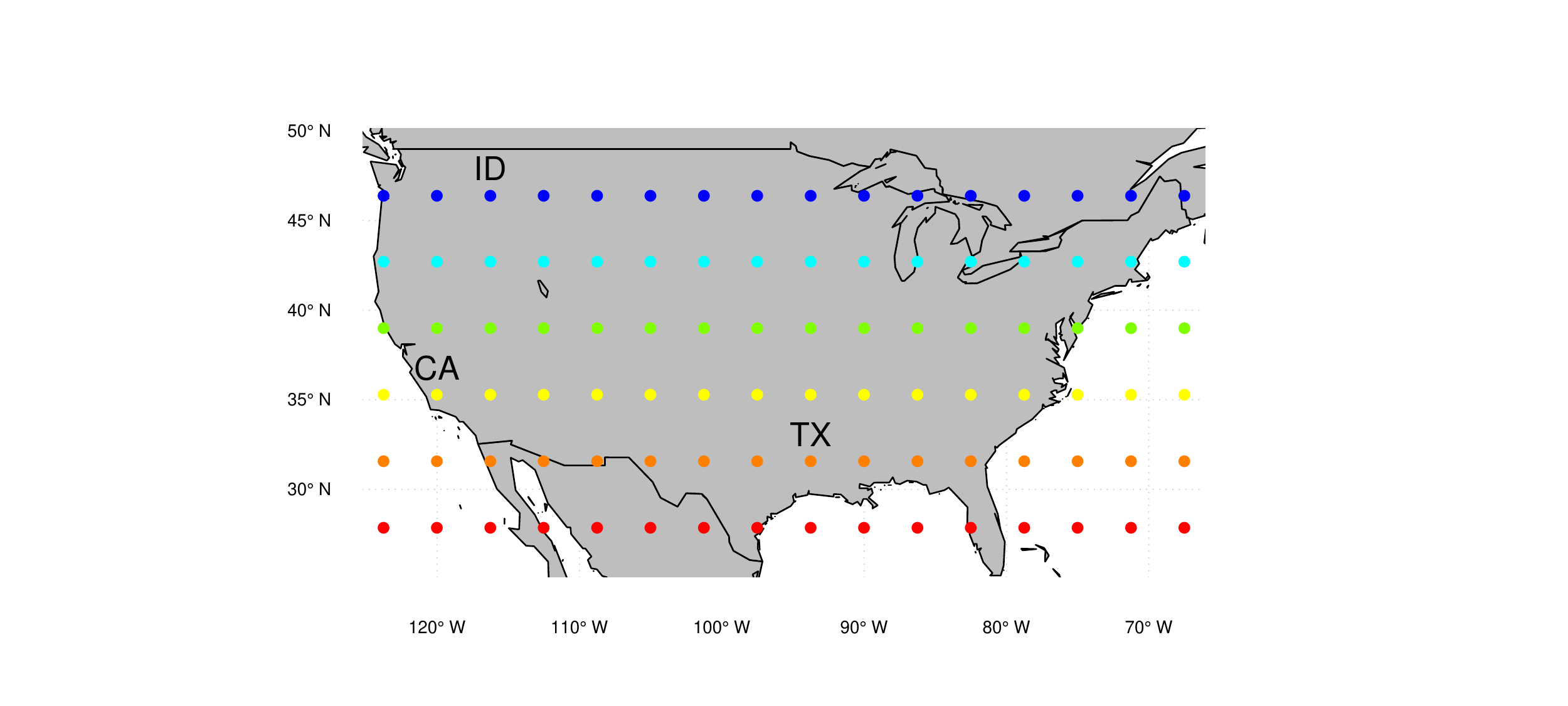}
\caption{Locations of centers of grid cells for model output used here. The rainbow color scale for latitudes is used throughout this study. Idaho (ID), California (CA), and Texas (TX) grid cells are used as examples in Figures \ref{fig:pixel_changes}, \ref{fig:rl_est_short_su}, and 
\ref{fig:rl_est_short}.}
\label{fig:locs}
\end{figure}

The GCM output used is part of an ensemble of climate
simulations completed by the Center for Robust Decision Making on Climate and Energy Policy (RDCEP) \citep[e.g.][]{castruccio2014,leeds2015}, using the Community Climate System Model version 3 (CCSM3) \citep{collins2006}, a fully-coupled model with full representation of the atmosphere (CAM3), land (CLM3), sea ice (CSIM5) and ocean (POP 1.4.3) components. The model was run at the relatively coarse T31 spatial resolution (3.75$^{\circ} \times$ 3.75$^{\circ}$ grid) \citep{yeager2006}, which made the lengthy runs used here possible. We use here the last 1000 years of output from each of three multimillennial runs with different atmospheric $\mathrm{CO}_2$ concentrations: 289 ppm (pre-industrial), 700 ppm (3.4 $^\circ$C increase in global mean temperature (GMT)), and 1400 ppm (6.1 $^\circ$C increase in GMT). In each case, solar forcing, aerosol concentrations, and 
concentrations of greenhouse gases other then $\mathrm{CO}_2$ are kept at their pre-industrial values. The final 1000 years of each model run are very nearly in equilibrium: net global radiative imbalances at the surface and at top-of-atmosphere (TOA) are smaller than 0.1 $\mathrm{W\,m^{-2}}$. The annual extreme temperatures therefore effectively form stationary time series. In this work, we consider annual maxima (from January 1 to December 31) of daily maximum temperature (T$_{\text{max}}$) and annual minima (from July 1 to June 31) of daily minimum temperature (T$_{\text{min}}$) over the contiguous United States and adjacent ocean regions (see Fig.~\ref{fig:locs}).

\section{Statistical background} \label{sec3}
The GEV distribution is widely applicable in the sciences because it arises, at least approximately, in many cases of natural data. 
The simplest situation where a GEV distribution can arise is in distributions of maxima taken from sequences of $n$ independent and identically distributed (i.i.d.) random variables ($Y_{1},\cdots,Y_{n}$), with ``block length'' $n$ sufficiently large. The Extremal Types Theorem says that if the maxima $M_{n}=\max (Y_{1}, \cdots, Y_{n})$, after normalization (i.e.\ $\frac{M_{n}-b_{b}}{a_{n}}$, $a_{n}>0$),   
converge in distribution as $n \rightarrow \infty$ (see Appendix~\ref{App:EVT}), then they converge to a GEV distribution \citep{fisher1928,gnedenko1943}. 

In this work, when studying high temperature extremes, the random variables $Y_{1,y},\cdots,Y_{n,y}$ are the daily maximum temperatures in year $y$, and the block length $n$ is 365 days, so that $M_{n,y}$ is the maximum for year $y$. Of course time series of daily temperatures are neither independent (they are somewhat autocorrelated) nor identically distributed (their distributions vary within a year). There is however theoretical justification for the use of GEV distributions in our case: it has been shown that the independence assumption can be relaxed for weakly dependent stationary time series \citep{leadbetter1983,hsing1991}, and \cite{einmahl2015} extended the theory to non-identically-distributed observations with conditions on the tail distributions (e.g.\ distributions share a common absolute maximum). 
In this work, we explicitly assess whether inferred GEV distributions actually provide an adequate description of the dataset in question. The CCSM3 temperature time series studied here do seem to meet this condition: quantile--quantile plots show that the GEV approximation fits reasonably well for most locations in our study area (see Fig.~\ref{fig:diag_su} and Fig.~\ref{fig:diag} in Appendix  \ref{App:diag}). We therefore assume that annual maxima and minima of daily temperatures in our model output can be approximated by GEV distributions. We will revisit the assessment of the GEV approximation in Section \ref{sec5} when we test whether our block length of one year is sufficient.

\subsection{GEV distributions} \label{gev}
We give here a brief review of EVT for block maxima. For further background, \cite{de2007} and \cite{resnick1987,resnick2007} give systematic theoretical accounts and \cite{coles2001} and \cite{beirlant2004} provide statistical treatments.  The GEV distribution function, described in terms of its three parameters $\mu$, $\sigma$, and $\xi$, is 
\begin{eqnarray} \label{GEV}
  G_{\mu, \sigma, \xi}(y) = \left\{ 
  \begin{array}{l l}
    \exp\left(-\left\{1+\frac{\xi(y-\mu)}{\sigma}\right\}_+^\frac{-1}{\xi}\right) & \quad  \xi \neq 0\\
    & \\
    \exp\left(-\exp\left\{\frac{-(y-\mu)}{\sigma}\right\}\right) & \quad \xi=0\\
  \end{array}\right.  
\end{eqnarray} 
where $u_{+} = \max(0,u)$. $\mu$ and $\sigma$ are location and scale parameters respectively, and the shape parameter $\xi$ determines the tail behavior of the density.   When $\xi$ is zero, the distribution is an \textit{exponentially--tailed} Gumbel distribution; $\xi>0$ yields the \textit{heavy--tailed} Fr\'{e}chet distribution; and $\xi <0$ yields the \textit{bounded-tailed} reversed Weibull distribution, in which the distribution has an absolute maximum of $\mu-\sigma/\xi$. Distributions of temperature extremes typically have $\xi<0$ \citep[e.g.][]{gilleland2006, fuentes2013}.

For studying low temperature extremes, we must consider minima rather than maxima. Equation \hyperref[gev]{(1)} can be used to study minima simply by considering that the maximum of $-Y_1,\ldots,-Y_n$, where $Y_{i}$ represents the daily minimum temperature on day $i$. However, if one lets $(\mu,\sigma,\xi)$ be the parameters for the 
corresponding GEV distribution of the maximum of the negative temperature, then increasing $\mu$ would correspond to decreasing temperature, which we find potentially confusing. We therefore adopt the convention that when considering temperature minima, for the corresponding GEV distribution, we write
\begin{eqnarray} \label{GEVmin}  G_{\mu, \sigma, \xi}(y) = \left\{
  \begin{array}{l l}
    \exp\left(-\left\{1+\frac{\xi(\mu-y)}{\sigma}\right\}_+^\frac{-1}{\xi}\right) & \quad  \xi \neq 0\\
    & \\
    \exp\left(-\exp\left\{\frac{-(\mu-y)}{\sigma}\right\}\right) & \quad \xi=0.\\  \end{array}\right.
\end{eqnarray}
With this definition, a larger $\mu$ corresponds to warmer temperature minima.
In the remainder of this work, we assume that for annual maxima, the relevant GEV distribution is \hyperref[gev]{(1)}, and for annual minima, the relevant distribution is \hyperref[GEVmin]{(2)}. 

\subsection{Relationship between GEV distributions and return levels} \label{sec3.2}

In the analysis that follows, we also describe extremes by their return levels, a widely used measure of extreme events. For warm temperature extremes, the $r$-year return level is the (block maximum) temperature exceeded 
on average once per $r$ years. Note that the concept of return levels implicitly assumes stationary conditions. Return levels in a dataset whose extremes follow a GEV distribution can be written in terms of the GEV parameters. Letting $p=1/r$, the $r$-year return level $y_{p}$ is the $100 \times (1-p)$th quantile of the underlying GEV distribution, which can be determined by solving the equation $G(y_{p})=1-p$ (for $0<p<1$) to obtain: 
\begin{equation} \label{quantile}  
y_{p} = \left\{
  \begin{array}{l l}
    \mu-\frac{\sigma}{\xi}\left\{1-\left(-\log\left(1-p\right)\right)^{-\xi}\right\} & \quad  \xi \neq 0\\
    \mu -\sigma \log\left(-\log\left(1-p\right)\right) & \quad \xi=0\\  \end{array}\right.
\end{equation} 
where the ``log'' denotes the natural logarithm. A similar result holds for temperature minima. \footnote{Note that \hyperref[quantile]{(3)} implicitly assumes block lengths of 1 year and must be changed for different block lengths: the definition of $p$ becomes $p=b/r$, where $b$ is the block length in years. The differences in $r$-year return levels defined by different block lengths are largest for the shortest return periods (largest $p$) and become negligible for multi-decadal return periods ($p$ close to 0).}

\begin{figure}[H]
\centering
\includegraphics[width=6.5in]{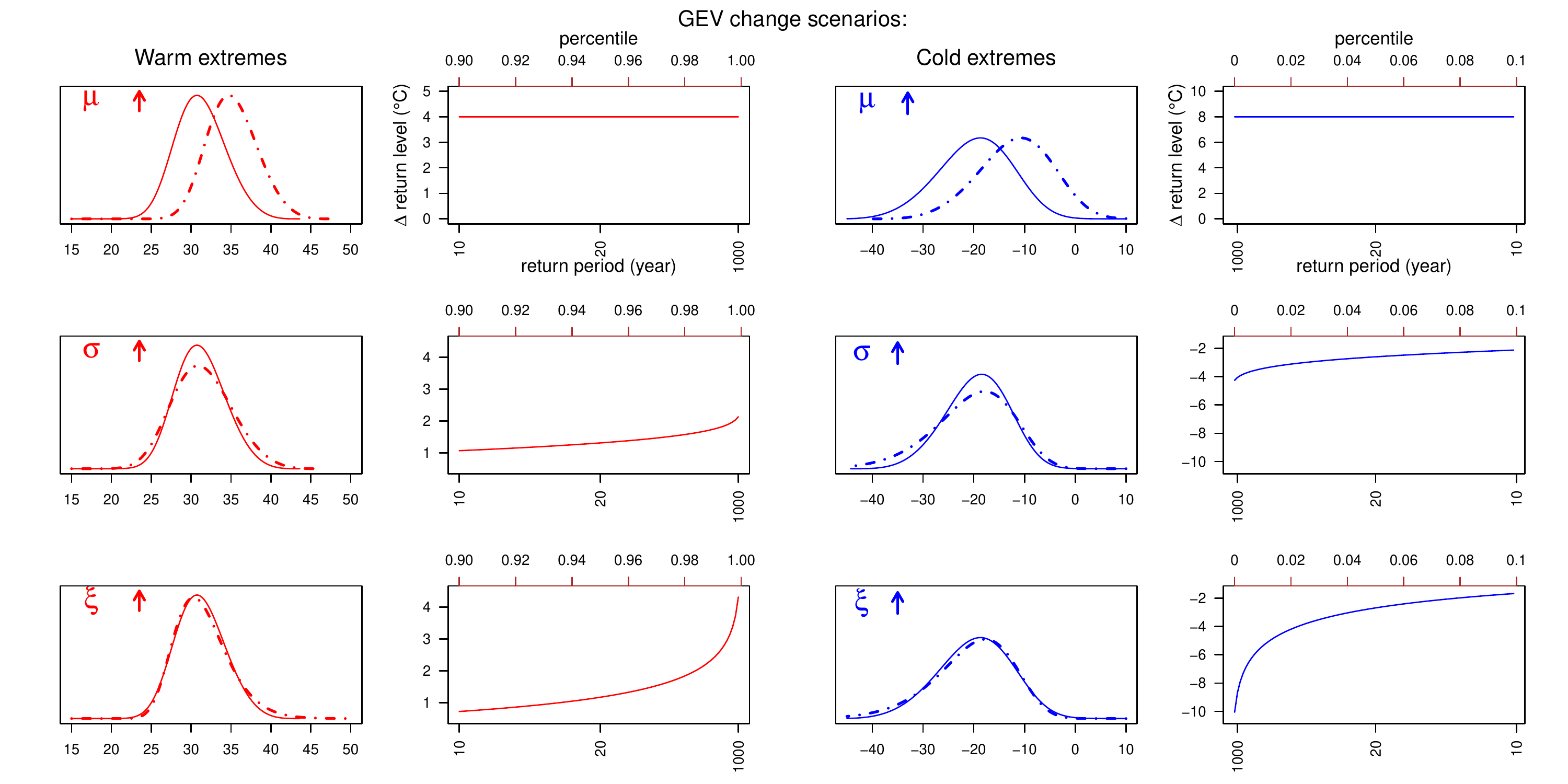}
\caption{Illustration of the effect on return levels of changing individual GEV parameters. We show consequences for both warm (red, left columns) and cold (blue, right columns) extremes. Columns 1 and 3 show GEV distributions for baseline (solid) and future (dashed) climates. Columns 2 and 4 show resulting changes in return levels for different return periods. Note that 1000-year periods are on the right for warm and left for cold extremes, to conform with percentiles. Location and shape parameters used here are chosen as representative of our model results, with larger effects in cold extremes than warm extremes, while shape parameter are identical in both cases. \textbf{Top row}: changing the location parameter shifts return levels uniformly across return periods.  \textbf{Middle row}: increasing the scale parameter produces effects dependent on return period. All return levels increase (decrease for cold extremes), but more so for longer return periods. \textbf{Bottom row}: increasing the shape parameter produces dramatic increases in return levels at very long return periods.}
\label{fig:gev_changes}
\end{figure}

In the warmer climate conditions that result from higher atmospheric $\mathrm{CO}_2$, the GEV distributions of temperature extremes may change, altering return levels. Changes in the different GEV parameters affect return levels in different ways.
In Figure~\ref{fig:gev_changes}, we illustrate the consequences on both distributions and return levels of changing each GEV parameter.
Changing the location parameter $\mu$ simply shifts the distribution of extremes by the change $\Delta \mu$ and changes return levels uniformly at all return periods by the same amount (Fig.~\ref{fig:gev_changes} top row). Changing the scale parameter $\sigma$ broadens or narrows the distribution of extremes, so that for warm extremes, increasing $\sigma$ increases return levels for long return periods and decreases them for short return periods (Fig.~\ref{fig:gev_changes} middle row). Changing the shape parameter $\xi$ alters the ``heaviness'' of the tails, but does so asymmetrically, with the greatest extremes most strongly affected (the high temperature tail for maxima, and the low temperature tail for minima). The resultant changes in return levels are then highly nonlinear with return period. Moderate changes in $\xi$ have negligible effect on return levels for periods less than 10 years, but strongly affect the ``extreme extremes" at long return periods.

\section{Results} \label{sec4}
By fitting the annual maxima and minima to a GEV distribution (see Appendix \ref{App:fitting} for details), we can identify changes in the distribution of extremes in possible future warmer climates, and evaluate how the characteristics of those changes alter return levels of extreme events. We show results here for all three model runs (pre-industrial, 700 ppm, and 1400 ppm $\mathrm{CO}_2$). 

\subsection{GEV parameters in preindustrial and future climates} \label{sec4.1}

\begin{figure}[H]
\centering
\includegraphics[width=6in]{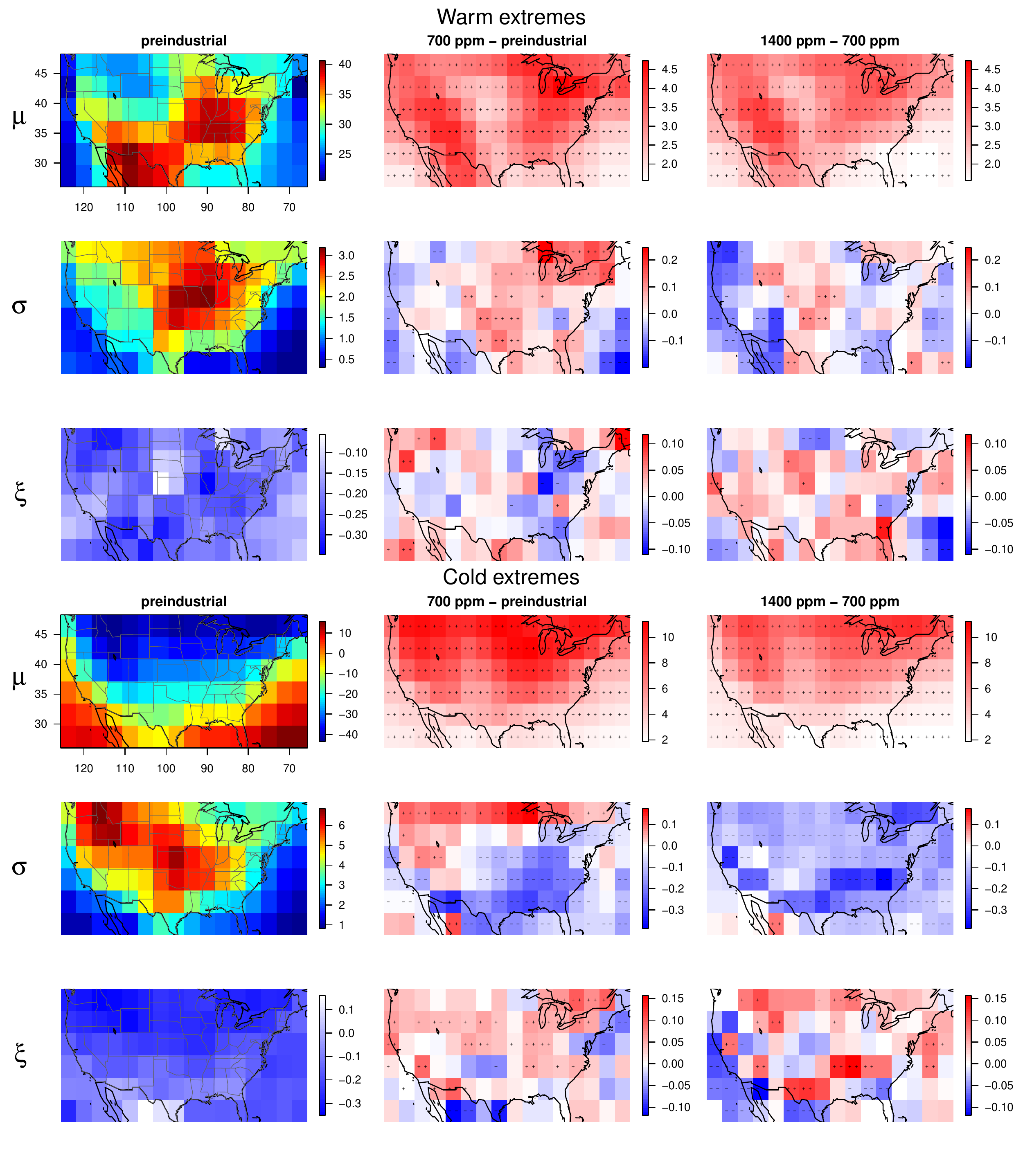}
\caption{The estimated CCSM3 GEV parameters and their changes in possible future climate conditions. \textbf{Left}: fitted GEV parameters (location $\mu$, scale $\sigma$, and shape $\xi$) for annual extremes for the baseline model run at pre-industrial CO$_2$ concentration. Top 3 panels are for warm extremes and bottom are for cold extremes. 
Negative $\xi$ is expected for temperature distributions.  \textbf{Middle}: changes in parameters ($\Delta \mu, \Delta \log \sigma , \Delta \xi$) after a warming of global mean temperature by 3.4 $^\circ$C (by raising atmospheric $\mathrm{CO}_2$ to 700 ppm). \textbf{Right}: changes in parameters after an additional 2.7 $^\circ$C warming (by raising $\mathrm{CO}_2$ from 700 to 1400 ppm). The symbols $++$ or $--$ mean the bootstrapped p-value for testing whether the parameter is different in the scenarios is $< 0.02$; $+$ or $-$ mean the p-value is between $0.02$ and $0.10$.}
\label{fig:par_changes}
\end{figure}

We show in Figure \ref{fig:par_changes} the estimated CCSM3 GEV parameters for warm and cold temperature extremes over the North American region. We show maps of GEV parameter values for the pre-industrial climate (left column), and maps of changes between pre-industrial and 700 ppm $\mathrm{CO}_2$ (middle column) and between 700 and 1400 ppm $\mathrm{CO}_2$ (right column). The fitted GEV parameters of the pre-industrial run show spatially coherent patterns, with both latitudinal gradients and land-ocean contrast that are consistent with expectations. 
The location parameters for warm extremes are highest where summers are warmest, in the desert Southwest and the inland Northeast/Midwest. Similarly, for cold extremes, $\mu$s are lowest where winters are coldest: their pattern shows a strong latitudinal gradient, but moderated along the coasts. 
The scale parameter, which affects the spread of the distribution of extremes, is largest in the continental interior and very small over the ocean, as expected. The shape parameter, which affects the far tail of the distribution of extremes, is negative nearly everywhere, as is usually the case for surface temperatures. 

Under warmer future climate conditions, we have strong reasons to expect positive shifts in location parameters: extremes should shift to warmer values for both warm and cold extremes. There are, however, no simple physical arguments that guide expectations for changes in the scale and shape parameters. The 1000-year model runs used here allow us to accurately estimate changes in all three parameters under this model.

\begin{figure}[H]
\centering
\includegraphics[width=5in]{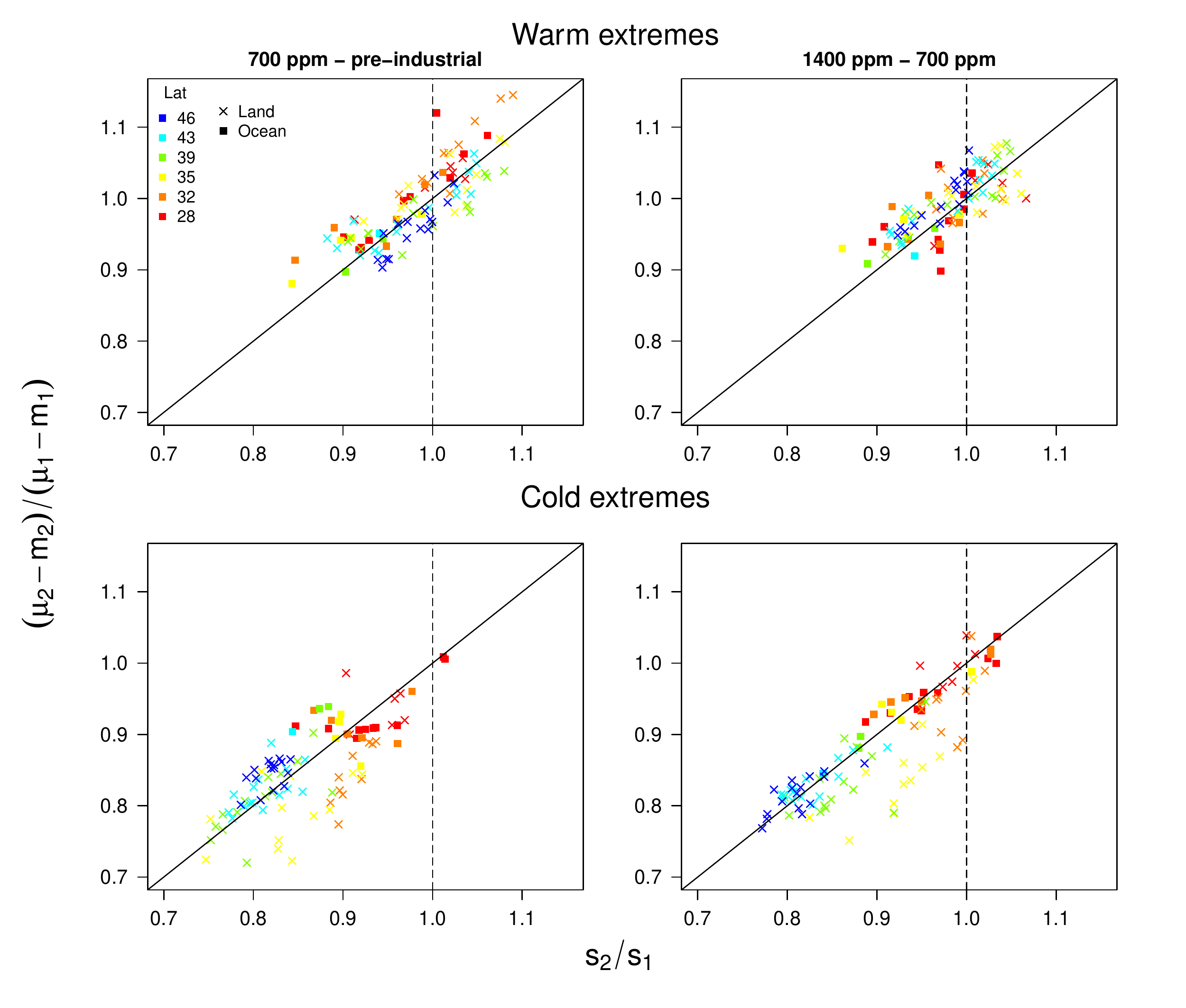}
\caption{Assessment of whether \textit{changes} in the location parameter of extremes are linked to changes in mean $(m)$ and standard deviation ($s$) of the corresponding seasonal distributions. \textbf{Left}: ratio of differences between GEV location parameter and seasonal mean compared to ratio of within-season standard deviations, for the transition from pre-industrial (climate state 1) to 700 ppm $\mathrm{CO}_2$ (climate state 2). \textbf{Right}: same results with climate state 1 corresponding to 700 ppm and state 2 to 1400 ppm. Scatter plots show these terms for all locations in study area, for warm extremes in upper panels and for cold extremes in lower panels. Plots show that $\mu_{2}-m_{2}$ is very well approximated by $(\mu_{1}-m_{1})s_{2}/s_{1}$, where $s_2/s_1$ is the ratio of within season standard deviations.}

\label{fig:gev_mu_sigma}
\end{figure}

Changes in the location parameters for both warm and cold extremes are, as expected, positive everywhere in the study region (Fig. \ref{fig:par_changes}, first and fourth rows) as climate warms.  As previously found in other studies of climate model output \citep{kharin2007, kharin2013} and observations \citep{lee2014}, the changes are unequal, with a larger amount of warming in cold extremes than in warm extremes. For warm extremes, location parameter changes are fairly
uniform spatially and similar to changes in summer mean temperatures (see Fig.~\ref{fig:ratio_gev_mu_su} in Appendix \ref{App:mean_extremes}). 
For cold extremes, location parameter changes show a strong latitudinal gradient, similar to the pattern in seasonal mean changes but with greater magnitude. That is, changes in cold extremes exceed changes in winter means (Fig.~\ref{fig:ratio_gev_mu_wi} in Appendix \ref{App:mean_extremes}). This asymmetry is readily explained by the asymmetric changes in the seasonal temperature distributions. In simulated warmer climate conditions, temperature variability (the standard deviation of the distribution) is relatively unchanged in summer but strongly reduced in winter, especially at higher latitudes \citep[e.g.][]{holmes2015}. Fig. \ref{fig:gev_mu_sigma} shows that for both warm and cold extremes, changes in the location parameter are well-explained by changes in means and standard deviations of seasonal temperature distributions.

For both warm and cold extremes, scale parameters show changes that are geographically complicated but statistically significant relative to pre-industrial values over much of the region (see Fig.~\ref{fig:par_changes}). For warm extremes, scale parameters show little change over much of the study area but modest increases over the eastern U.S., up to 20\% larger than pre-industrial values. Scale parameters change more strongly for cold extremes, with a strong latitudinal gradient over land. The change from pre-industrial to 700 ppm CO$_2$ leads to statistically significant decreases in scale parameters at lower latitudes but statistically significant increases at higher latitudes. Scale parameters for cold extremes additionally show clear nonlinear effects (see Fig. \ref{fig:par_changes} fifth row, middle and right panels): scale parameters increase at high latitude locations in the transition from pre-industrial conditions to 700 ppm, but then decrease at these same locations in the transition from 700--1400 ppm, partially negating the earlier increases. Shape parameters show few statistically significant changes for warm extremes but significant positive shifts for cold extremes in some areas. When considering the full change in climate states from pre-industrial to 1400 ppm CO$_2$, changes in the shape parameter for cold extremes are in general significant at most inland locations, indicating that linear transformations are not adequate for explaining all changes in extremes (see Fig.~\ref{fig:delta_xi_289_1400}).

\subsection{Changes in return levels} \label{sec4.2}
\begin{figure}[H]
\centering
\includegraphics[width=6.5in]{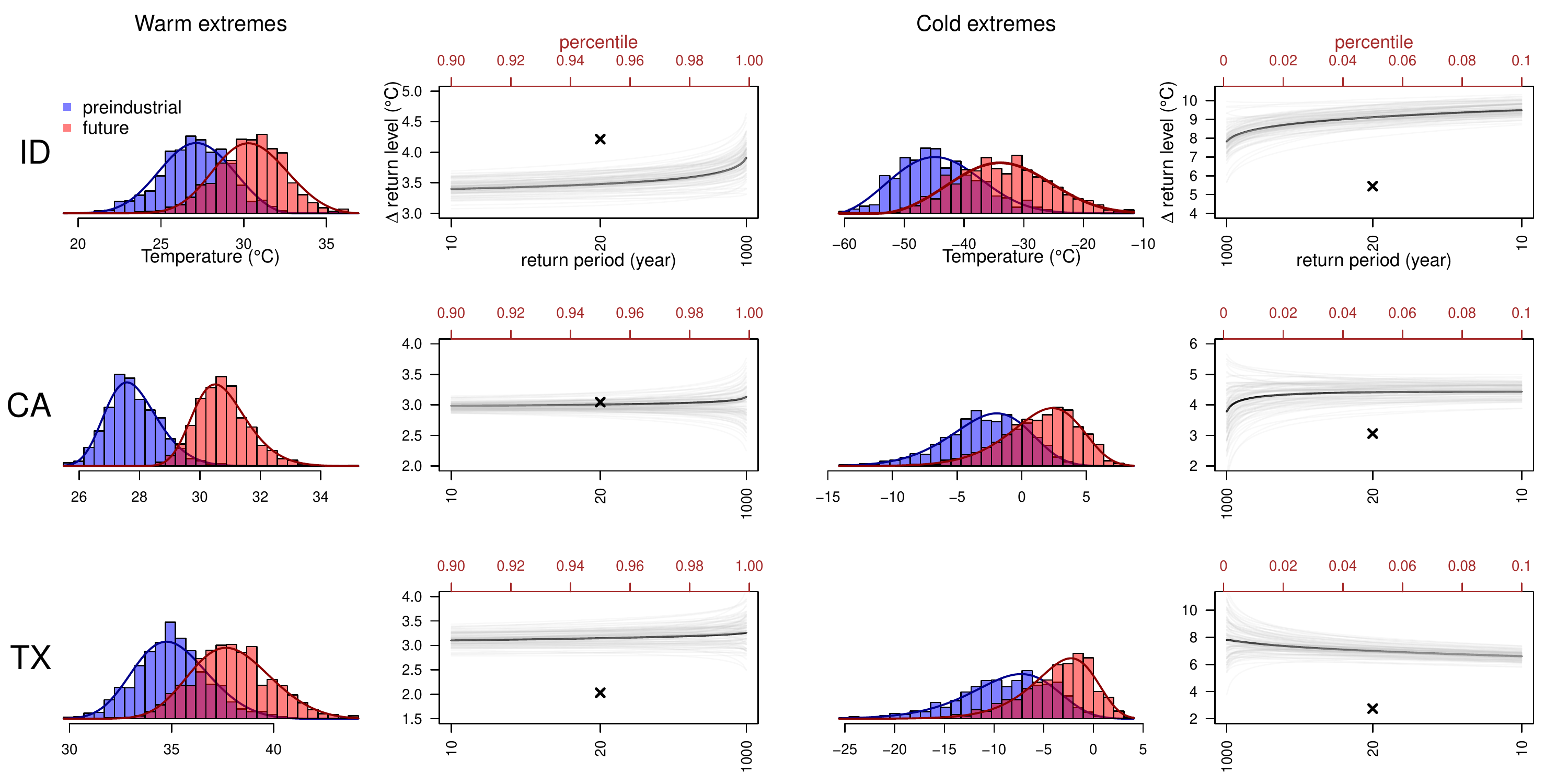}\\
\caption{Illustration of the changes in return levels and their relationship with the changes in seasonal means. The histograms are for the 1000 annual maxima (warm extremes) and minima (cold extremes) for preindustrial (289 ppm, blue) and future (700 ppm, red).  Smooth curves are corresponding estimated GEV distributions. Return level plots show changes in estimated return levels (dark curve) and uncertainty from the bootstrap procedure (lighter curves). The corresponding changes in seasonal means are marked with a cross (change in mean summertime daily maximum for warm extremes and mean wintertime daily minimum for cold extremes).}
\label{fig:pixel_changes}
\end{figure}

As discussed in Section~\ref{sec3.2}, it can be helpful to describe changes in tail behavior in terms of changes in return levels. We showed in Figure \ref{fig:gev_changes} how hypothetical changes in individual GEV parameters affect return levels at different return periods. Here we present examples of changes in the actual fitted GEV distributions of model output, and show how these produce changes in return levels. We use as examples three model locations (individual grid cells), located in Idaho (ID), California (CA), and Texas (TX) (see Fig.~\ref{fig:locs} for locations), that illustrate the behavior discussed in Section \ref{sec4.1}. For warm extremes in all three locations (Fig.~\ref{fig:pixel_changes}, left columns), the changes in return levels are roughly constant for all return periods (except for ID at return periods approaching 1000 years), meaning the dominant change is simply a shift in the distribution of extremes, i.e.\ a change in location parameter. The shifts are close (within 1$^\circ$C) to the shifts in mean seasonal temperatures, as discussed previously. For cold extremes (Fig.~\ref{fig:pixel_changes}, right columns), in the inland locations of TX and ID the shifts in extremes are much larger than in the seasonal means and the scale and shape factors play a 
significant role, producing return level changes that can vary with return period (though differently in different locations). Of the examples shown here, scale and shape parameters play the least significant role in the one coastal location (CA), as would be expected.

\begin{table}[H]
\caption{GEV parameter estimates of warm and cold extremes for pre-industrial climate state and the estimated changes from pre-industrial to 700 ppm climate state at ID, CA, and TX locations. ``**'' means the
bootstrapped p-value for testing whether the parameter is different is $< 0.02$, ``*'' means
the p-value is between $0.02$ and $0.1$.}
\resizebox{\textwidth}{!}{%
\begin{tabular}{cccclllccclll}
\hline
   & \multicolumn{6}{c}{Warm extremes} & \multicolumn{6}{c}{Cold extremes} \\ \cline{3-6} \cline{9-12}
  & \multicolumn{3}{c}{289 ppm} & \multicolumn{3}{c}{700-289 ppm} & \multicolumn{3}{c}{289 ppm } & \multicolumn{3}{c}{700-289ppm}\\
  & $\hat{\mu}$ & $\hat{\sigma}$ & $\hat{\xi}$ & $\Delta\hat{\mu}$ & $\Delta\hat{\sigma}$ & $\Delta\hat{\xi}$ 
  & $\hat{\mu}$ & $\hat{\sigma}$ & $\hat{\xi}$ & $\Delta\hat{\mu}$ & $\Delta\hat{\sigma}$ & $\Delta\hat{\xi}$ \\ 
   
\hline
\hline
ID   & 26.4    & 2.1  & -0.32   & +3.3 ** & -0.03 & +0.05 *&  -41.9   & 7.2  & -0.37   & +11.0 ** & +0.7 **& +0.03 \\
CA   & 27.5  & 0.8   &  -0.15  & +2.9 ** & +0.02  &  +0.01  & -1.5  & 2.9 & -0.14 & +4.2 **& -0.3 ** & +0.04 *   \\
TX   & 34.4  & 1.7  & -0.21  & +2.9 ** & +0.25 ** &  -0.01   & -6.9  & 4.1 & -0.10 & +4.8 **& -1.1  **& +0.05     \\
\hline
\end{tabular}%
}
\label{table:1}
\end{table}

\begin{figure}[H]
\centering
\includegraphics[width=6in]{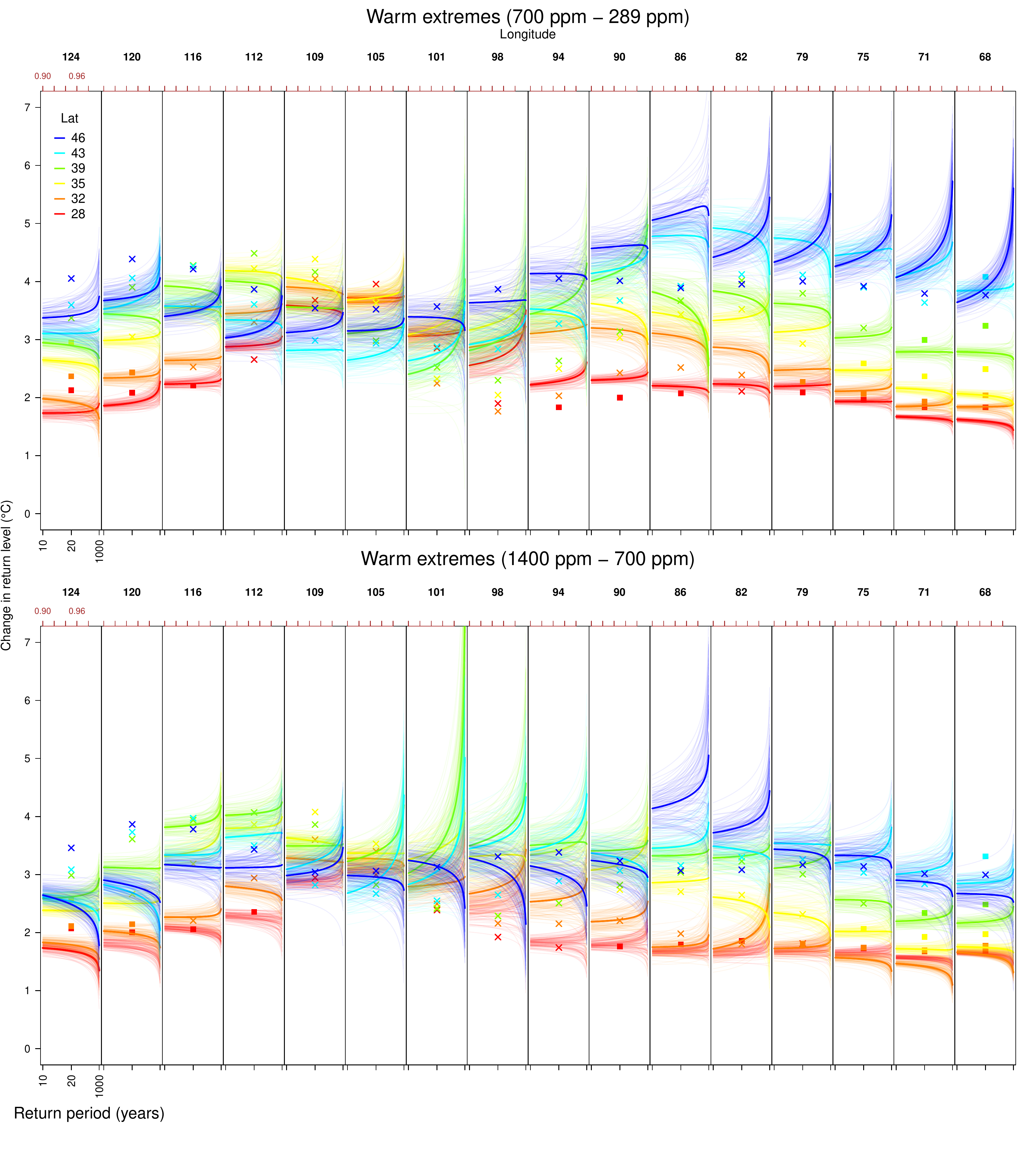}\\
\caption{Estimated changes in warm extremes return levels for possible future climate conditions in our CCSM3 runs. \textbf{Upper}: 700 ppm $\mathrm{CO}_2$ vs. 289 ppm $\mathrm{CO}_2$ climate conditions, \textbf{Lower}: 1400 ppm $\mathrm{CO}_2$ vs. 700 ppm $\mathrm{CO}_2$ climate conditions. Each panel shows results from the 6 grid cells in a longitude band, while each color represents a latitude band. Estimated changes in return levels are plotted as solid lines in their corresponding colors. Estimates obtained from block bootstrapped samples by resampling years are plotted as lighter and thinner curves to form envelopes representing the associated uncertainties. The corresponding summertime mean changes for inland and ocean grid cells are marked by cross and filled box symbols, respectively.}
\label{fig:rl_changes_su}
\end{figure}

\begin{figure}[H]
\centering

\includegraphics[width=6in]{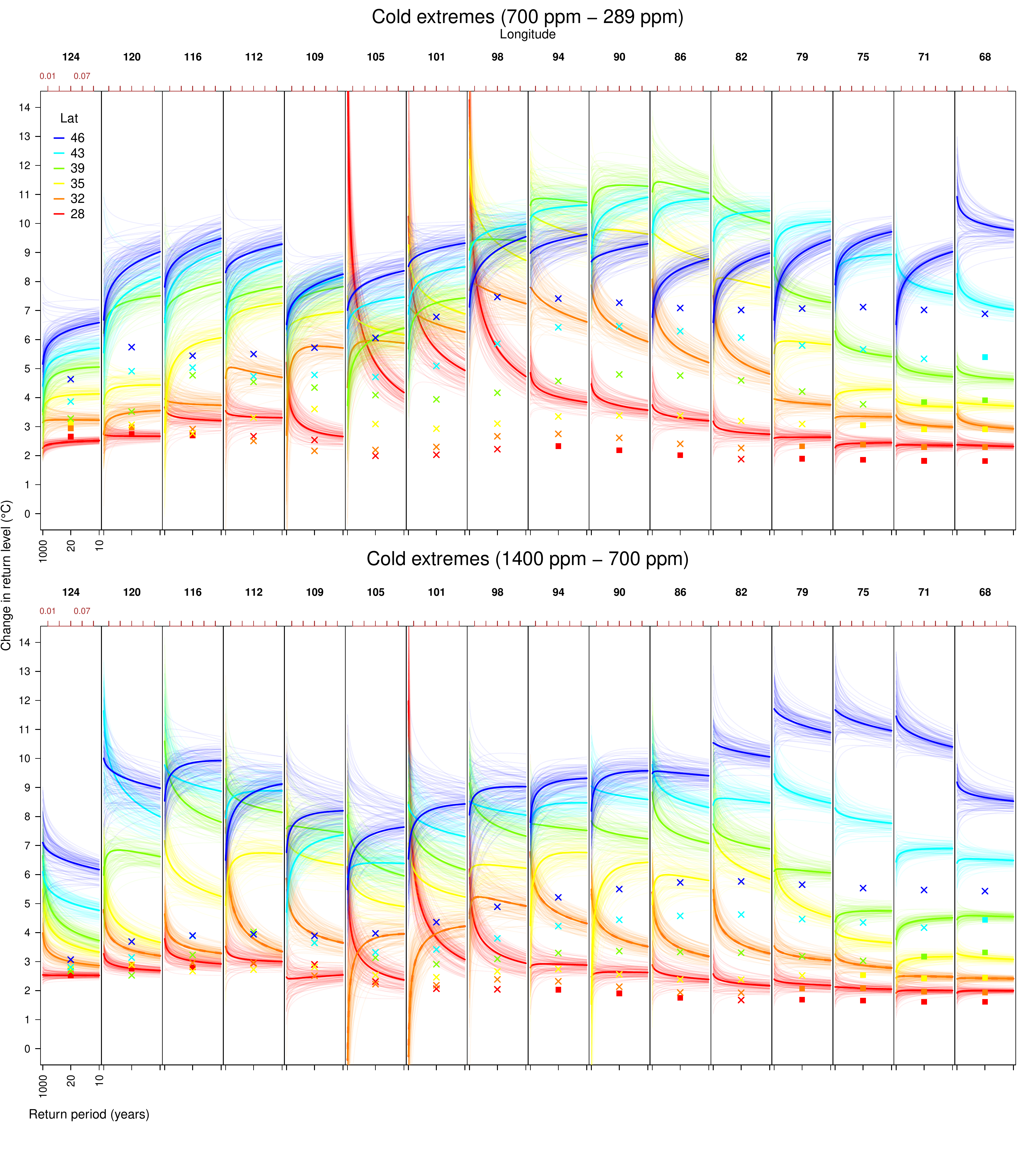}\\
\caption{As in Fig. \ref{fig:rl_changes_su} but for annual temperature minima. Crosses and boxes now give wintertime mean changes. Note that the return period axis is now flipped so that longer return periods are on the left. In general, cold extremes warm more than do winter means, especially at high latitudes.} 
\label{fig:rl_changes_wi}

\end{figure}

The different changes in warm and cold extremes shown in the example locations of Figure 6 are characteristic of the whole contiguous United States. In CCSM3 model output, throughout the region, annual maximum return level changes follow changes in summer means, but annual minimum return level changes exceed winter means, with stronger influence of the scale and shape parameters. Figures~\ref{fig:rl_changes_su} and \ref{fig:rl_changes_wi} show the changes in return levels across return periods from 10 to 1000 years, for annual maxima and minima, respectively. For warm extremes, changes in return levels are relatively flat in both ocean and inland locations, implying that the dominant effect is a change in the location parameter. Exceptions include a small part of the central Midwest corn belt region and the far Northeast, which show increases in return levels with return periods. The sharply rising ``extreme extremes'' in a few Midwest locations are due to the combined increases in both the scale and the shape parameter; these effects grow still further in the transition to a 1400 ppm climate (Fig.~\ref{fig:rl_changes_su} lower panel). For cold extremes (Fig.~\ref{fig:rl_changes_wi}), return level changes for ocean locations are flat, but inland locations tend to show striking effects due to changes in scale and shape parameters. Effects differ by latitude: at low latitudes, return levels tend to increase with longer return periods, sometimes dramatically so, whereas at high latitudes, the pattern generally reverses, so that the most extreme cold temperatures tend to increase less than modestly extreme cold temperatures. This distinction is particularly apparent in the transition from pre-industrial to 700 ppm CO$_2$.

\section{Sensitivity analysis: GEV block size} \label{sec5}
In this analysis, as in many climate applications of EVT, we have assumed that an annual block is long enough that the GEV distribution is approximately valid. Our long time series allow us to explicitly evaluate this assumption. The evaluation relies on the fact that  the GEV distribution has the property of ``max-stability''.  In the context of distributions of block maxima ($M_{n}$), max-stability means that the distribution after normalization is identical for any block size larger than $n$.

In our case, if an annual block is long enough that the GEV distribution is approximately valid for a given time series, then the shape parameter estimate $\hat{\xi}_n$ obtained with annual blocks should be approximately the same as that obtained with a larger block size $n'$ ($\hat{\xi}_{n'}$). The validity of the GEV approximation using annual extremes can therefore be assessed by checking the consistency of estimated shape parameters using longer block sizes (i.e.\ multiple years). 

\begin{figure}[H] 
\centering
\includegraphics[width=6.5in]{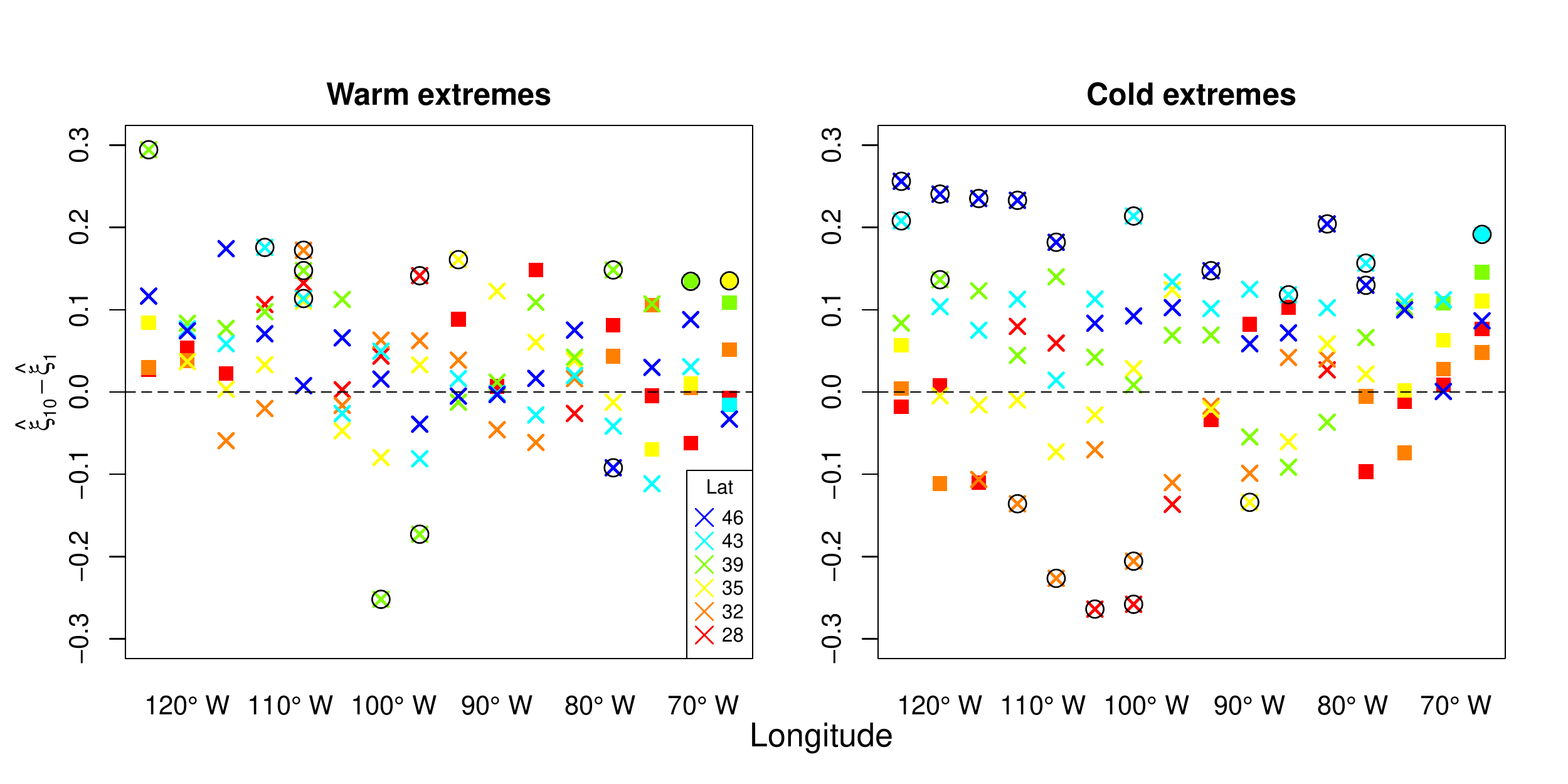}
\caption{A statistical diagnostic of the assumption that annual blocks are sufficiently long for the GEV approximation. Differences (10- vs.\ 1-year blocks) in estimated shape parameters are plotted for warm (\textbf{left}) and cold (\textbf{right}) extremes against longitude. Estimates are shown for the pre-industrial climate state, for all model grid cell locations in the study area, with the same color and symbol scheme as in previous figures. Subscripts in notation denote the block length, e.g.\ $\hat{\xi}_{\text{10}}$ means 10-year blocks. Circled points are those locations where differences in estimates of the shape parameter changes are significant at the 0.05 level based on their bootstrapped p-values.}
\label{fig:xi_diff}
\end{figure}

We refit GEV distributions for each grid cell in our study region with block maxima/minima sizes of 2, 5 and 10 years, and compare shape parameters. We find that for annual maxima, the differences in $\xi$ with block size are distributed around zero, suggesting that an annual block size may be sufficient (Fig. \ref{fig:xi_diff}, left panel). For annual minima, we see effects of block size depending on latitude: longer blocks produce larger $\xi$ at high latitudes, where changes in extremes are largest (Fig. \ref{fig:xi_diff}, right panel). These results suggest that one year may not be sufficiently long for the GEV approximation to be accurate. These results also suggest that quantile--quantile plots (see Fig.~\ref{fig:diag} in Appendix \ref{App:diag}) may not be sufficient to check for the appropriateness of the GEV distribution for the purpose of predicting events with longer return period than the data length. We observe however that annual blocks may be long enough for the study of the \textit{changes} of extremes, since changes in return levels are fairly consistent with block length (See Figs.~\ref{fig:diag_rl_su} and \ref{fig:diag_rl_wi}).

\section{Sensitivity analysis: data length on return level estimation}\label{sec6}
The thousand-year model runs used in this work provide fairly accurate estimates of changes in return levels even for long return periods. Estimated uncertainties on return level changes -- the bootstrapped envelopes in Fig.~\ref{fig:rl_changes_wi} -- are generally within 20\% of the estimated changes even for 100-year return periods.
This dataset therefore allows us to study empirically how well
GEV methods work when applied to model output at the shorter lengths (decades to centuries) more commonly used in climate studies. 

\begin{figure}[H]
\centering
\includegraphics[width=6.5in]{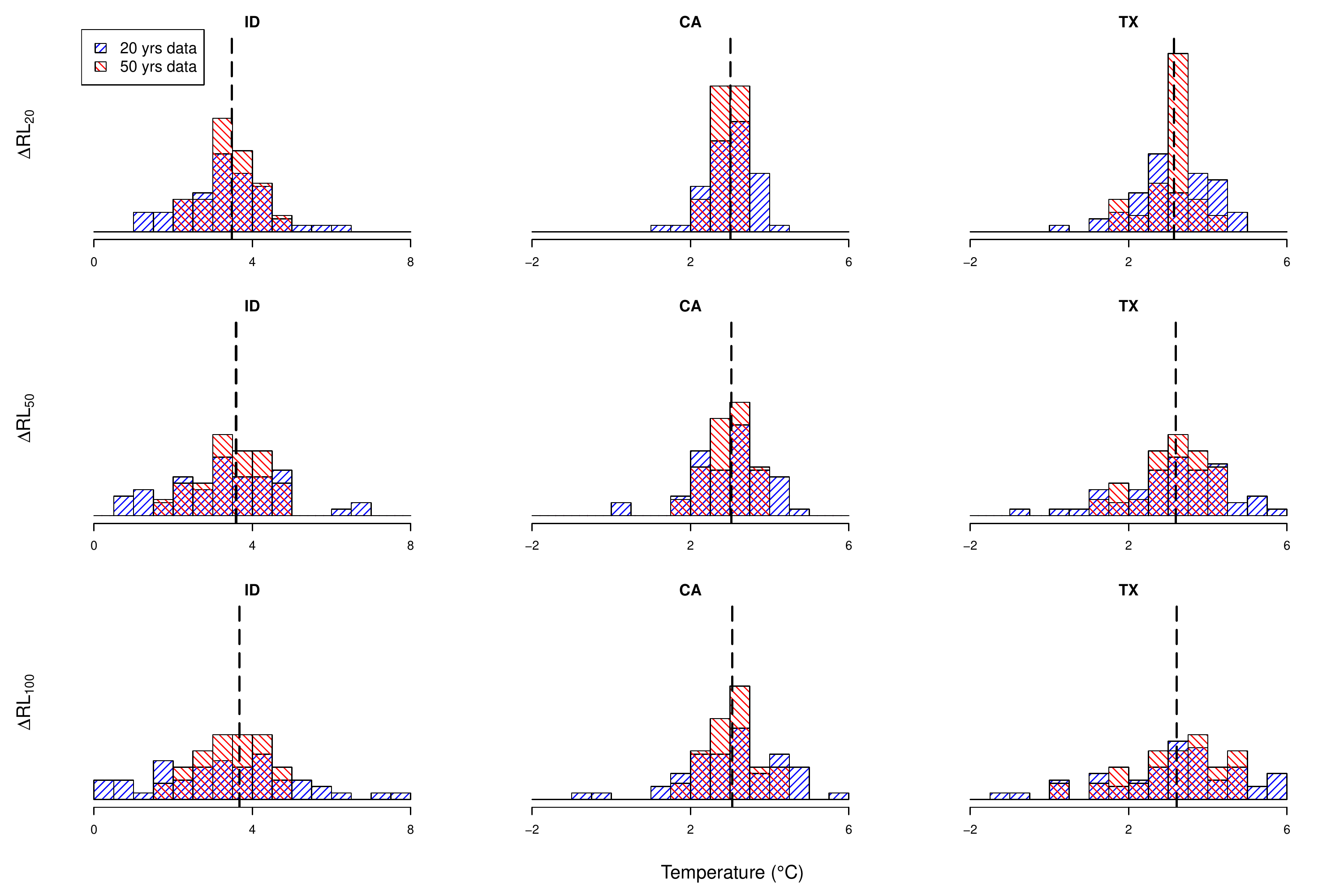}\\
\caption{The relationship between sample errors for return levels and data lengths for warm extremes. Estimates of changes (from pre-industrial to 700 ppm $\mathrm{CO}_2$) in return levels (20, 50, 100 year return periods) obtained by using different length segments (20 and 50 years, shown as blue and red histograms), as compared to the ``ground-truth'' changes obtained with the full 1000-year runs (dashed lines). We show warm extremes for the same three grid cells shown in previous figures, located in Idaho, California, and Texas. Pre-industrial return levels for these locations (for 20, 50, and 100 year periods) are ID: 30.4, 31.1, and 31.5 $^\circ$C; CA: 29.4, 29.8, and 30.1 $^\circ$C, and TX: 38.1, 38.9, and 39.4 $^\circ$C.}
\label{fig:rl_est_short_su}
\end{figure}

\begin{figure}[H]
\centering
\includegraphics[width=6.5in]{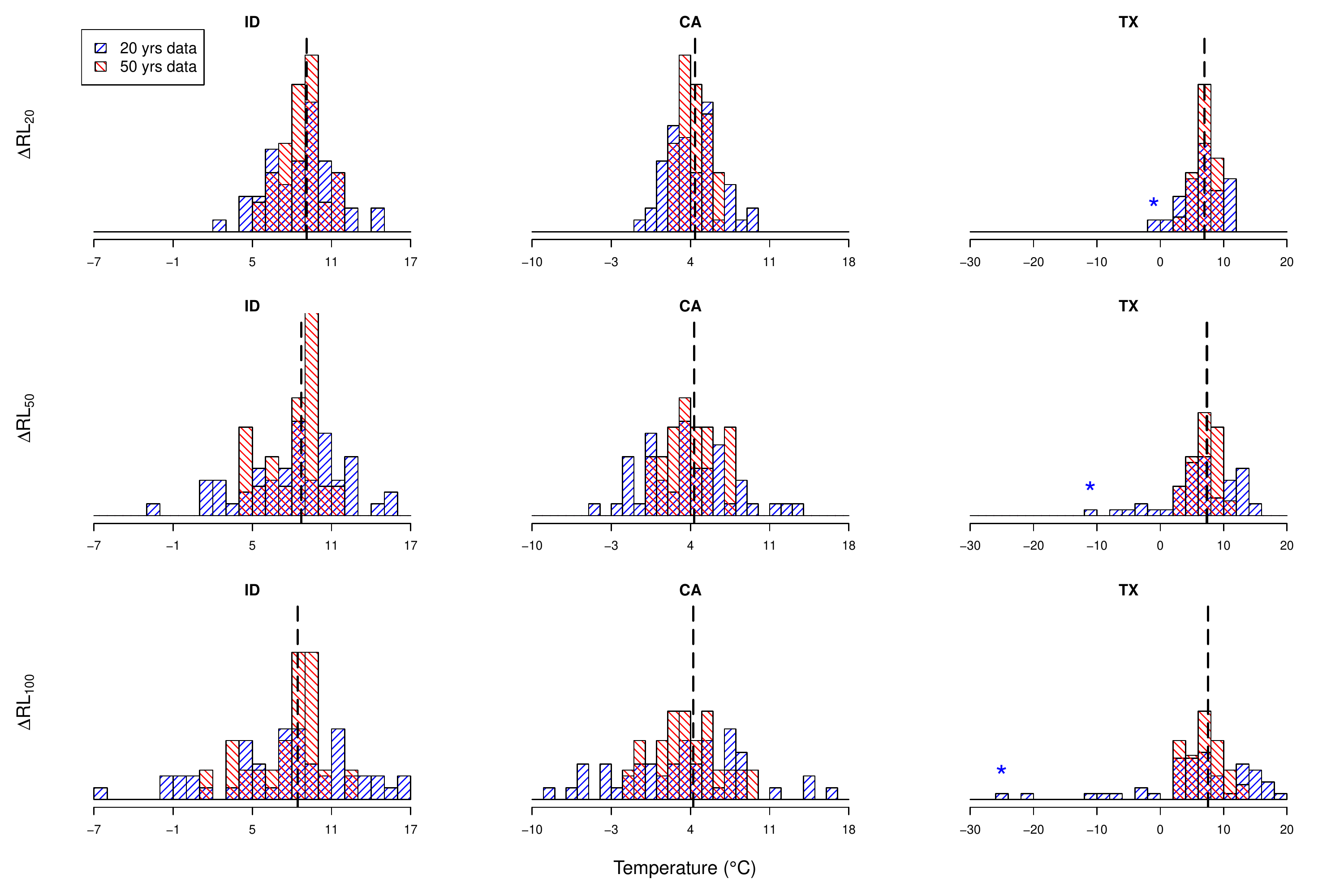}\\
\caption{Same as Fig. \ref{fig:rl_est_short_su} but for cold extremes. Pre-industrial return levels for these locations are ID: -54.9, -56.8, and -57.8 $^\circ$C; CA: -8.5, -10.1, and -11.2 $^\circ$C, and TX: -17.2, -19.9, and -21.8 $^\circ$C. Especially unreliable estimates of changes in return levels were obtained for one pair of 20-year segments in TX (denoted by *). This error is mainly due to a poorly estimated shape parameter: for this pair of 20-years segments, the estimated shape increased from $0.10$ to $0.60$ (the estimates given by PWM are $0.12$ and $0.11$) while the ``true'' values (based on 1000-year runs) were -0.10 and -0.05 for the preindustrial and 700 ppm scenarios, respectively.}
\label{fig:rl_est_short}
\end{figure}

To assess how uncertainties increase with shorter model runs, we divide the preindustrial and 700 ppm time series into segments of 20 or 50 years and refit the GEV parameters for each pair of segments (so, for example, pairing the first 20 years of the preindustrial run with the first 20 years of the 700 ppm run). The resulting distributions of estimates of changes in return levels for warm and cold extremes are shown in Figures \ref{fig:rl_est_short_su} and \ref{fig:rl_est_short}.

The results show that sampling error can be large, as expected, when using climate simulations of length comparable to the return periods of interest. In both warm and cold extremes, the distribution of estimates of return level changes derived from short model segments are centered around their ``true'' values but with large spread (Fig. \ref{fig:rl_est_short_su} and \ref{fig:rl_est_short}). In these examples, the sampling errors are comparable to their ``true'' changes in return level when 20-year segments are used. Unsurprisingly, spreads are largest when using short model segments to predict long return periods. For example, for the Idaho test location, when changes in annual minimum 100-year return levels are estimated using 20-year segments (Fig. \ref{fig:rl_est_short} lower left panel), those estimates can differ by $\pm 10\,^{\circ}\mathrm{C}$ while the ``true'' change is only $8.4\,^{\circ}\mathrm{C}$. Uncertainties can be affected by the choice of estimation method. \cite{hosking1985} suggests that estimation based on the method of probability weighted moments (PWM) has better small sample properties than maximum likelihood, as we use here. Fig~\ref{fig:pwm} shows that result are generally very similar for the two methods, although PWM does avoid the few cases of highly unrealistic estimates seen in Figure~\ref{fig:rl_est_short} in the TX example location.

\begin{figure}[H]
\centering
\includegraphics[width=5.75in]{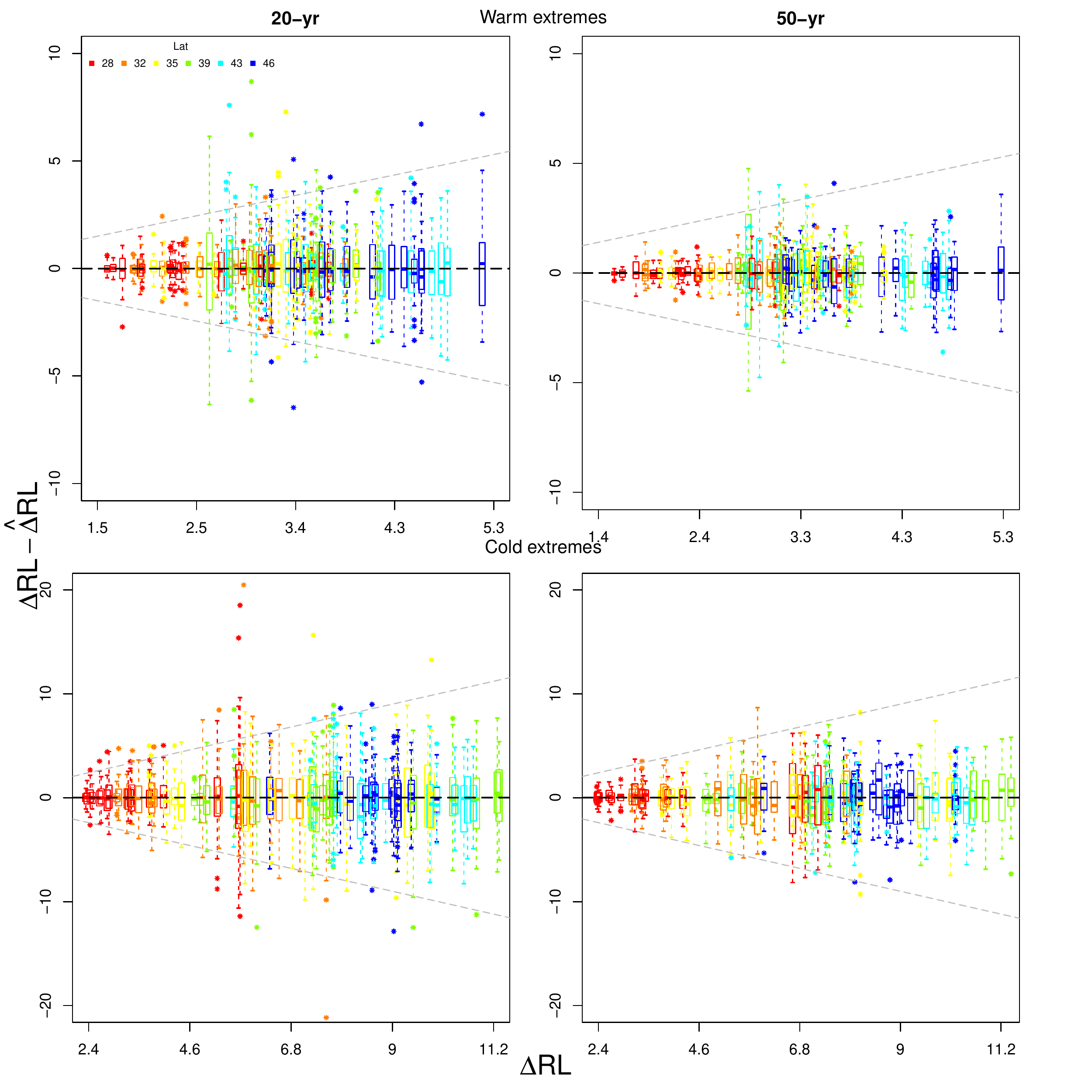}\\
\caption{Assessment of sampling errors in estimates of return level changes using comparable data lengths for all model grid cells in the study area. Values shown are the estimated return level change minus the ``ground-truth'' change determined from the entire 1000-year segment ($\Delta \text{RL}-\hat{\Delta}\text{RL}$), plotted against the ``ground-truth'' change ($\Delta \text{RL}$). \textbf{Left}: distribution (displayed as a boxplot) of estimates of changes in 20-year return levels made using 20-year segments. \textbf{Right}: the same for 50-year return levels and 50-year segments. Top and bottom rows show warm and cold extremes. Gray reference lines are the 1:1 lines (i.e. sampling error
$=$ return level change).}
\label{fig:rl_50_50}
\end{figure}

The example locations shown in Figures \ref{fig:rl_est_short_su} and \ref{fig:rl_est_short} are representative of land locations in the entire study area. In Figure \ref{fig:rl_50_50}, we show the spread in estimates for all model grid cells. We show two cases: estimating changes in 20-year returns (both warm and cold extremes) with 20-year segments and changes in 50-year returns with 50-year segments. The length of the model segment is especially important for temperature extremes, because their distribution has a bounded tail. For inland locations, the spread in estimates is comparable to the true change for the 20-year case but somewhat lower for the 50-year case. Ocean locations show much smaller sampling error, presumably because changes in extremes are produced predominantly by shifts in the location parameter and not the scale or shape parameters, which are more difficult to estimate. The locations with the largest uncertainties are indeed those with the largest changes in shape parameter: a few Midwest locations for warm extremes and some locations in Mexico for cold extremes.

It is important to note that spatial modeling approaches may be able to reduce estimation variability. (See further discussion in following section.) When not using such approaches, this assessment suggests that when using single model runs or observational data, sampling error may be quite large when the length of the series is comparable to the return period of interest.

\section{Discussion} \label{sec7}
In this study, we have used extreme value theory to study changing temperature extremes in model projections of future climate states. Following prior work, we assume the GEV model provides a reasonable approximation for the distribution of annual temperature extremes (see Fig \ref{fig:diag_su} and Fig \ref{fig:diag}) and that changing temperature extremes can therefore be studied by estimating changes in GEV parameters and resulting implied changes in return levels.  We use millennial-scale equilibrated simulations of pre-industrial and 700 and 1400 ppm $\mathrm{CO}_2$ concentrations (producing changes in global mean temperature of 3.4 and 6.1$^\circ$C) to explore both the physics of climate model projections and the utility of statistical approaches based on GEV distributions in different contexts.

For our model runs, the results suggest that for the contiguous United States, much of the behavior of temperature extremes in higher CO$_2$ climate states  appears a straightforward consequence of changes in the mean and standard deviation of the underlying temperature distributions. Annual warm extremes generally shift simply in accordance with mean shifts in summer temperatures. Annual cold extremes warm more strongly than do winter mean temperatures, but largely do so as expected given decreased variability in wintertime temperatures. In both cases, the changes in the location parameter of annual extremes appear well-explained by shifts in overall seasonal temperature means and standard deviations (Fig. \ref{fig:gev_mu_sigma}). In most inland locations in the winter, changes in the scale and shape parameters can provide an additional complication, producing substantial differences in return level changes at longer return periods (Fig.~\ref{fig:rl_changes_wi}). Note that these results are from a single climate model, and there is no guarantee that any model captures all aspects of changes in extremes well \citep[e.g.][]{parey2010}. Model studies are, however, important given the difficulty of evaluating changes in extremes from the short observational record.

The millennial-scale model runs used here allow us to assess the validity of the use of annual block sizes in studies of GEV distributions of temperature extremes. Our results suggest that annual blocks are sufficiently long for representing warm extremes, but may be insufficient for cold extremes, especially for inland locations at high latitudes. In these locations, altering the block length alters the shape parameter of the estimated GEV distribution (Fig. \ref{fig:xi_diff}). In general, increasing block size reduces biases, but can increase sampling error in estimates if the time series is limited. The choice of block size therefore requires consideration of inherent trade-offs. 

Millennial scale model runs also enable detailed investigation of the relationship between data length and GEV sampling error. Previous studies have suggested that extrapolation should be carried out with caution \citep[e.g.][]{kharin2007}. Our long runs allow us to better quantify estimation uncertainties resulting from short models runs. We find that over inland locations, estimation uncertainty is indeed large when using series length is comparable to the return period of interest (Fig. \ref{fig:rl_50_50}). Ocean locations show much smaller sampling error, in large part because temperature variability is much lower over oceans than over land. If using very short series (e.g.\ 20 years), lack of data mandates care in choosing the appropriate estimation procedure \citep{hosking1985, gilleland2006}. Our 1000-year model runs allow us to concretely demonstrate the large uncertainties in estimates of changes in extremes that result when time series length is comparable to the return period of interest.

The computational demands of millennial-scale climate simulations restricts us here to examining a single climate model run at fairly coarse resolution. A single model should not be taken as robust guidance on how temperature extremes may change in future climate conditions. However, few modeling groups have performed runs of comparable length, and the multi-model data in public archives are not ideal for estimating changes in extremes: run lengths are much shorter ($\approx$ 100 years), the number of realizations of any scenario is small, and climate conditions in the simulations are evolving rather than stationary. Estimating GEV distributions that are changing over time is also more challenging than in the equilibrium setting, making the limitations of short model runs even more pertinent. If understanding changes in extremes in a changing climate is a research priority, larger ensembles of model runs would be helpful. 

In the absence of large ensembles, one could potentially overcome some of the challenges of limited data by exploiting the spatial structure of temperature extremes (see Fig. \ref{fig:par_changes}) to ``borrow strength over space''. Numerous prior studies do explicitly model the spatial structure of a climate variable of interest, e.g.\ \cite{cooley2007, cooley2010, craigmile2013, wang2016}. However, spatial modeling is not ideal: it may introduce bias, it greatly complicates the computation, and spatial structure in GEV parameters can be difficult to distinguish from spatial dependence in observed extremes. We do not use these approaches in this work since the millennial-scale model runs allow us to estimate the GEV parameters and their changes accurately by fitting them to each model grid cell separately. However, where only shorter time series (decades to centuries) are available, such as in analyses of observations, appropriate spatial modeling may be essential to reduce estimation uncertainty. 

Extreme value theory (EVT) has become popular for studying climate extremes, but our results here suggest that one should be aware of the underlying assumptions, the corresponding implications, and the potential limitations when applying it. EVT using block extremes can potentially allow characterization of tail behavior that differs from that of the underlying distribution, but it involves a corresponding penalty, as fitting a distribution of block extremes necessarily involves throwing out most of one's data. For short time series, care must be taken to ensure that the drawbacks do not outweigh the benefits.  In the climate model output studied here, for example, shifts in the location parameter for both warm and cold extremes are well-explained by changes in the mean and standard deviation of the underlying temperature distribution. The millennial time series used here allow us to also identify significant changes in the scale parameter, especially for cold extremes, but with shorter time series, sampling errors can be too large for estimated return levels to be of much practical value. Long model runs such as those used here therefore provide an important tool for the study of climate extremes, helping both in clarifying the contexts in which EVT should best be used and in devising approaches for working with shorter time series.

\section*{Acknowledgements} \label{acknowledgements}
This work was conducted as part of the \href{https://www.statmos.washington.edu}{Research Network for Statistical Methods for Atmospheric and Oceanic Sciences} (STATMOS), supported by NSF Award \#s 1106862, 1106974, and 1107046, and the \href{http://www.rdcep.org/}{Center for Robust Decision Making on Climate and Energy Policy} (RDCEP), supported by the NSF Decision Making Under Uncertainty program Award \# 0951576.

\begin{appendices}
\renewcommand{\thesection}{A.\arabic{section}}
\renewcommand{\thefigure}{A.\arabic{figure}}

\section{Comparison of distribution of extremes with overall distribution: summer} \label{App:ID_su}
We show the comparison of the overall distribution and the distribution of extremes, as in Fig.~\ref{fig:ID_wi} but here for summer rather than winter. 

\setcounter{figure}{0}  
\begin{figure}[H] 
\centering
\includegraphics[width=4in]{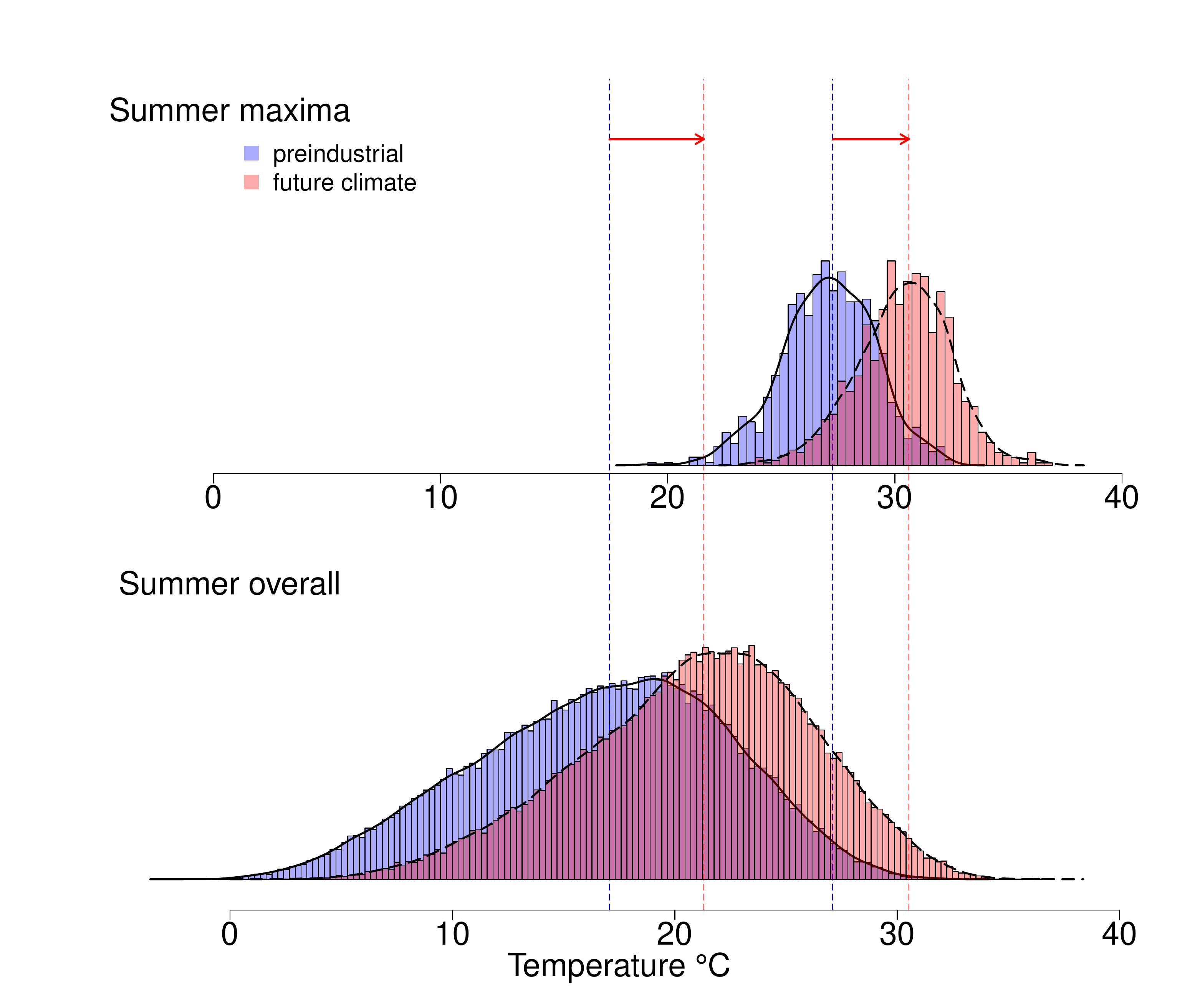}
\caption{As in Figure \ref{fig:ID_wi} but for summer warm extremes. \textbf{Lower}: overall distribution of summer (JJA) daily maximum temperature (T$_{\text{max}}$) in a north Idaho location, for baseline and 700 ppm CO$_2$ climate states. \textbf{Upper}: distributions of seasonal maxima of summer T$_{\text{max}}$. Note that the median shift of the overall distribution is now slightly larger than that of the distribution of extremes; the opposite was true for winter cold extremes.}
\label{fig:ID_su}
\end{figure}

\section{Extremal types theorem} \label{App:EVT}
In this appendix, we briefly describe the fundamental result in extreme value theory that justifies the use of GEV model for block maxima. 
Let $\{Y_{i}\}$ be a sequence of independent and identically distributed random variables with cumulative distribution function $F$ and let $M_{n}=\max\{Y_{1}, \cdots. Y_{n}\}$. When $n \to \infty$, the sequence of random variables $\left\{M_{n}\right\}$ converges to a single point $y_{F}=\sup\{y: F(y)<1\}$, where $y_{F}$ maybe infinite. 
In order to have a useful description of the distribution of $M_{n}$ when $n$ is sufficiently large, normalization is needed, that is, we are seeking a limiting distribution for $\frac{M_{n}-b_{n}}{a_{n}}$. It turns out that if there exist constants $a_{n}>0$ and $b_{n}$ and a non-degenerate distribution function $G$ such that
$$\mathbb{P}\left(\frac{M_{n}-b_{n}}{a_{n}} \le y\right) \stackrel{d}{\rightarrow} G(y)$$
then $G$ must be of the same type (two random variables $X$ and $Y$ are of the same type if $aX+b$ has the same distribution as $Y$) as one of the three extreme value classes below:
\begin{align*}
\textit{Gumbel: } G(y)&=\mathrm{exp}(-\mathrm{exp}(-y)) \qquad -\infty<y<\infty;\\
\textit{Fr\'{e}chet: } G(y)&=\left\{ 
  \begin{array}{l l}
    0 & \quad y \leq 0,\\
    \mathrm{exp}(-y^{-\alpha}) & \quad  y>0, \qquad \alpha>0;
  \end{array} \right.\\
\textit{reversed Weibull: } G(y)&=\left\{ 
  \begin{array}{l l}
    \mathrm{exp}(-(-y)^{\alpha}) & \quad y < 0, \qquad \alpha >0,\\
    1 & \quad  y \geq 0;\\
  \end{array} \right. 
\end{align*}
where $\alpha = 1/\xi$ for Fr\'{e}chet and $\alpha = -1/\xi$ for reversed Weibull (see \hyperref[GEV]{(1)} and \hyperref[GEVmin]{(2)}). Conversely, any distribution function of the same type as one of these extreme value classes can appear as such a limit. The GEV distribution in \hyperref[GEV]{(1)} provides a unifying representation of these three types of distributions. 

\section{GEV diagnostics} \label{App:diag}
\begin{figure}[H]
\centering
\includegraphics[width=5in]{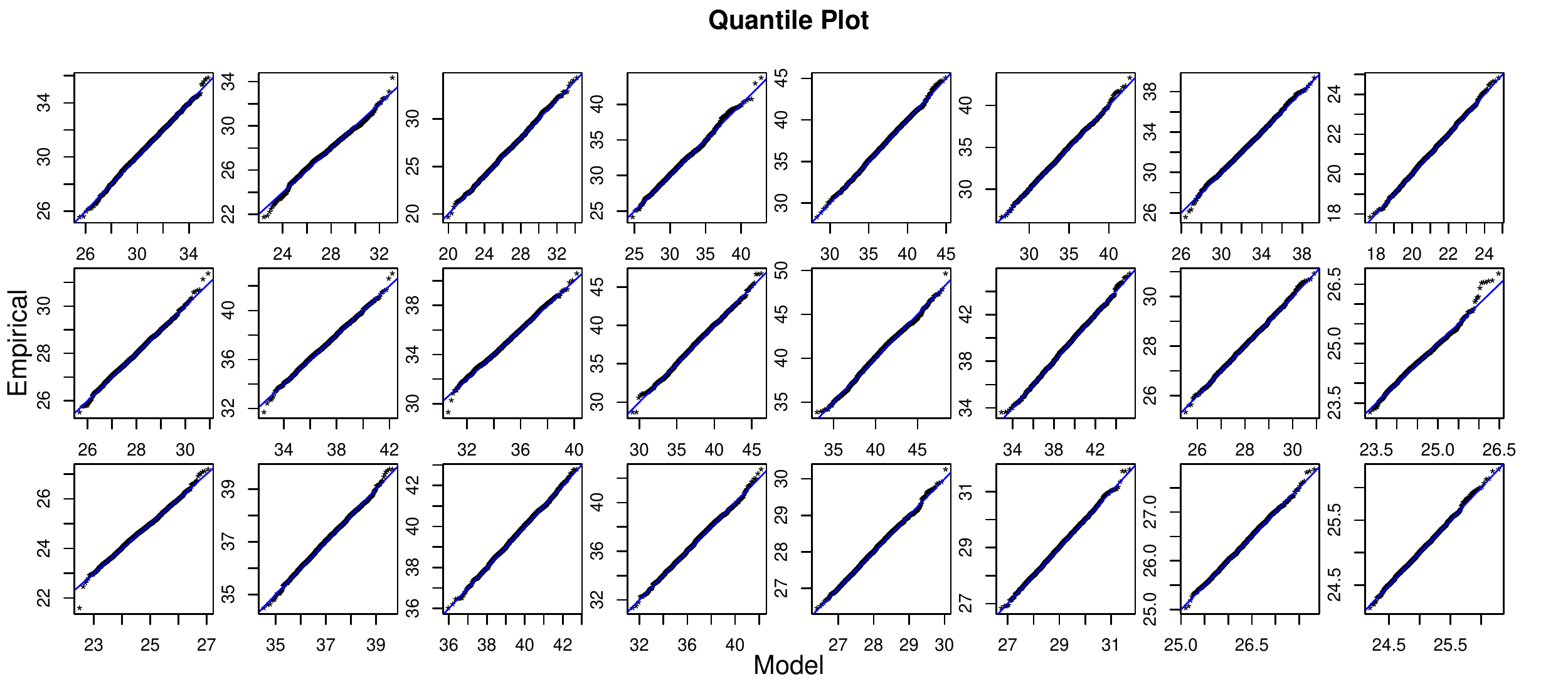}
\caption{The Q-Q plots for the fitted GEVs of warm extremes in pre-industrial climate. For clarity we display every other row and column of grid cells in Fig \ref{fig:locs}.}
\label{fig:diag_su}
\end{figure}

\begin{figure}[H]
\centering
\includegraphics[width=5in]{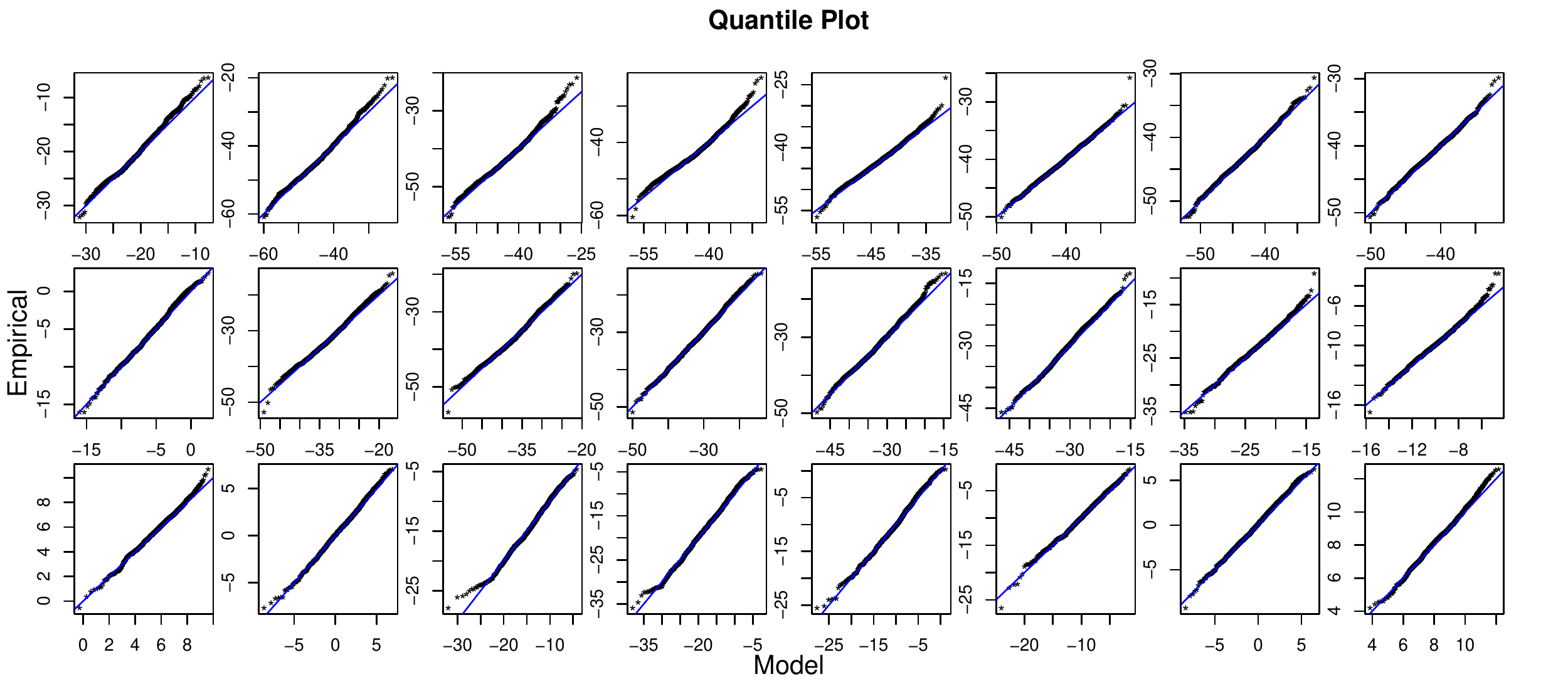}
\caption{As in Fig. \ref{fig:diag_su} but for cold extremes.}
\label{fig:diag}
\end{figure}

We use quantile-quantile plots to assess the goodness of fit of GEV distributions to model seasonal temperature extremes in the study region. For most (but not all) of the locations, there is a good agreement between empirical quantiles and the fitted quantiles. This agreement holds across climate states for both warm and cold extremes.  (See Figs.~\ref{fig:diag_su} and \ref{fig:diag}.)

\begin{figure}[H]
\centering
\includegraphics[width=5in]{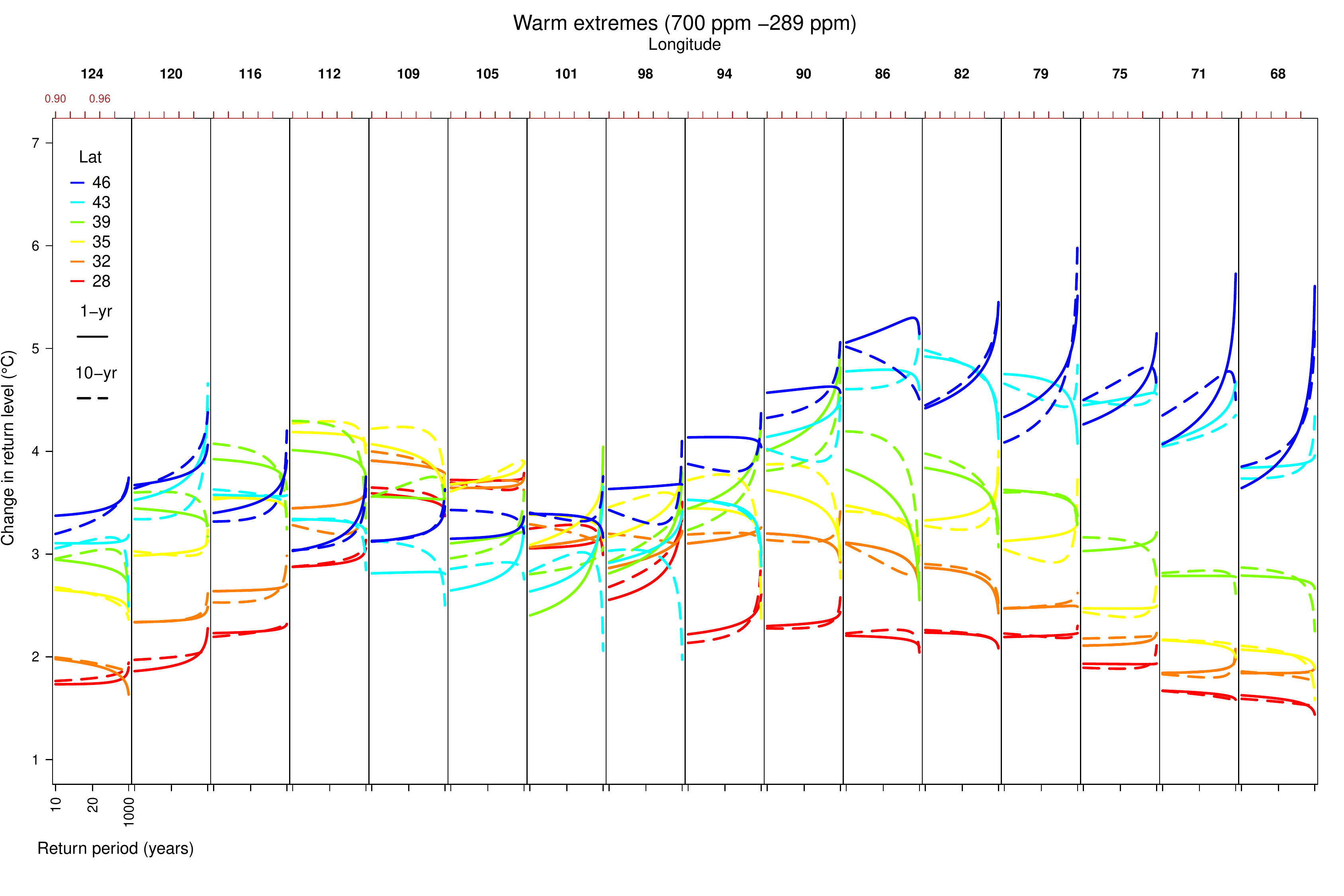}
\caption{As in Fig. \ref{fig:rl_changes_su} (upper panel) but here comparing the estimated changes obtained from using annual maxima (solid lines) versus decadal maxima (dashed lines).}
\label{fig:diag_rl_su}
\end{figure}

\begin{figure}[H]
\centering
\includegraphics[width=5in]{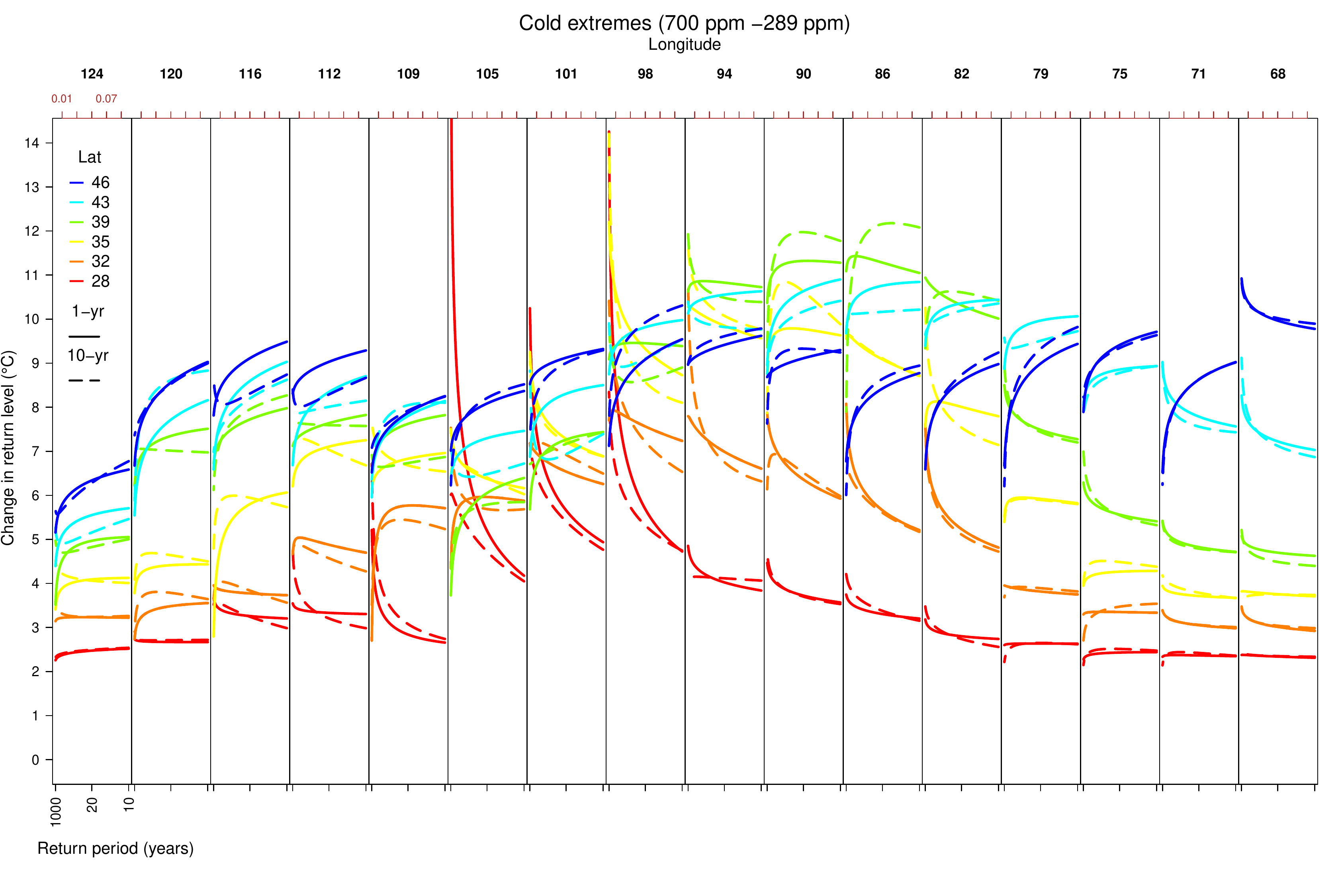}
\caption{As in Fig. \ref{fig:rl_changes_wi} (upper panel) but here comparing the estimated changes obtained from using annual minima (solid lines) versus decadal minima (dashed lines).}
\label{fig:diag_rl_wi}
\end{figure}

In Section \ref{sec5}, we showed that, based on the max-stable property of GEV, annual blocks may not be sufficiently long for GEV distribution to be valid for the purpose of extrapolating more extreme events (i.e. return periods are longer than the data length). Here we investigate whether annual blocks are appropriate to study the \textit{changes} in return levels by comparing the results obtained from annual blocks versus decadal blocks (See Figs.~\ref{fig:diag_rl_su} and \ref{fig:diag_rl_wi}.)  To infer return levels for annual extremes from decadal extremes, we need to make the additional assumption that annual extremes are independent across years. 

\section{Model fitting procedures} \label{App:fitting}
The parameters of the extreme value distributions were fitted using maximum likelihood 
to the annual maxima/minima of each grid cell in the study region. 
The numerical optimization needed to find these estimates were performed by using the function \texttt{gev.fit} in the \textsf{R} package \texttt{ismev} \citep{ismev}. Return levels were obtained by plugging in 
the estimated GEV parameters into (\ref{quantile}). 

\begin{figure}[H] 
\centering
\includegraphics[width=5.5in]{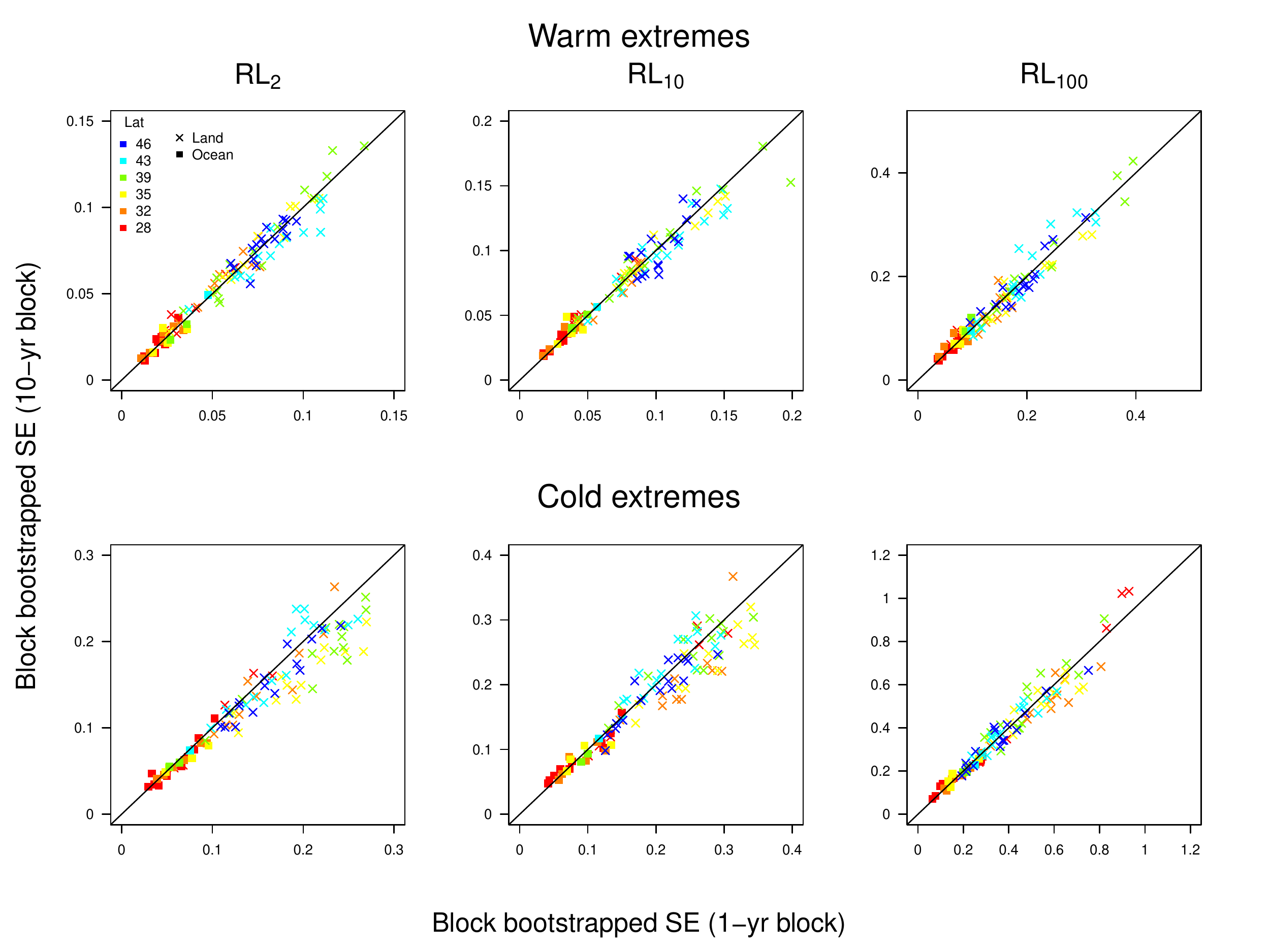}\\
\caption{Assessment of estimation uncertainty to return levels using different block sizes for block bootstrap. The scatterplot shows block bootstrapped standard errors by resampling years (x-axis) and decades (y-axis) of 2-year (left), 10-year, and 100-year (right) return levels.}
\label{fig:block_boot}
\end{figure}

We assess the uncertainties for GEV parameters and return levels by using bootstrap resampling. Both simple nonparametric bootstrap \citep{efron1979} and circular block bootstrap \citep{politis1992} were applied to annual extremes. Specifically, let $m_{i,j}$ be the annual extreme for year $i$ and site $j$ and $\mathbf{m}_{i}=\{m_{i,1}, \cdots, m_{i,k}\}$ where $k$ is the total number of sites. $\{\mathbf{m}_{i}\}_{i=1}^{n}$ is split into $n$ ($n=1000$ years in this study) overlapping temporal blocks of length $b$ ($b$ =1,2,5,10 years): years $1$ to $b$ will be block $1$ (i.e.\ $\{\mathbf{m}_{i}\}_{i=1}^{b}$), years $2$ to $b+1$ will be block $2$ and block $1000$ is years $1000,1,\cdots,b-1$. From these $n$ blocks, $n/b$ blocks are drawn at random with replacement. Then aligning these $n/b$ blocks will give the bootstrap observations. This approach allows us to provide plausible uncertainties for our estimates without having to model spatial dependencies or temporal dependencies for annual extremes within blocks; standard errors for different block sizes show little differences suggesting that extremes from one year to the next are nearly independent (see Fig.~\ref{fig:block_boot}). 

\section{Relationships between changes in mean and changes in extremes} \label{App:mean_extremes}
\begin{figure}[H]
\centering
\includegraphics[width=5in]{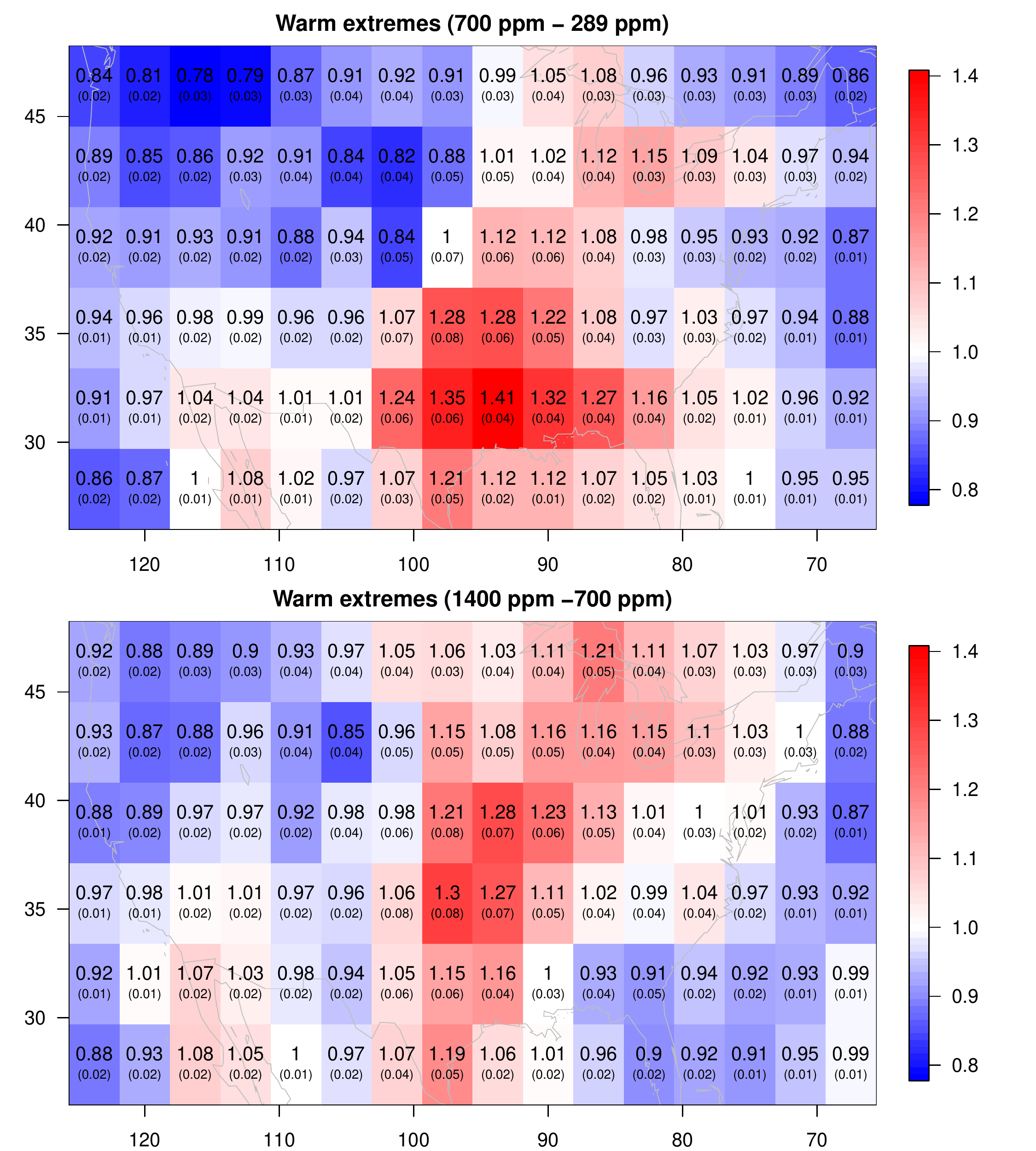}
\caption{An investigation of the relationship of changes in the location parameter of warm extremes (i.e. $\Delta \mu$) to changes in summer means ($\Delta m$). \textbf{Upper}: the ratio of changes in the location parameter of extremes to changes in means for the climate state from pre-industrial to 700 ppm. \textbf{Lower}: the ratio for the climate states 700 ppm and 1400 ppm. Red (blue) grid cell location indicates the shift of the GEV distribution is larger (smaller) than the shift of the overall distribution. Numbers on the top are the estimated ratios (i.e. $\Delta \hat{\mu}/\Delta m$) and numbers in parentheses are the bootstrapped standard errors for $\Delta \mu$. These ratios are generally near 1 in the contiguous United States.}
\label{fig:ratio_gev_mu_su}
\end{figure}

\begin{figure}[H]
\centering
\includegraphics[width=5in]{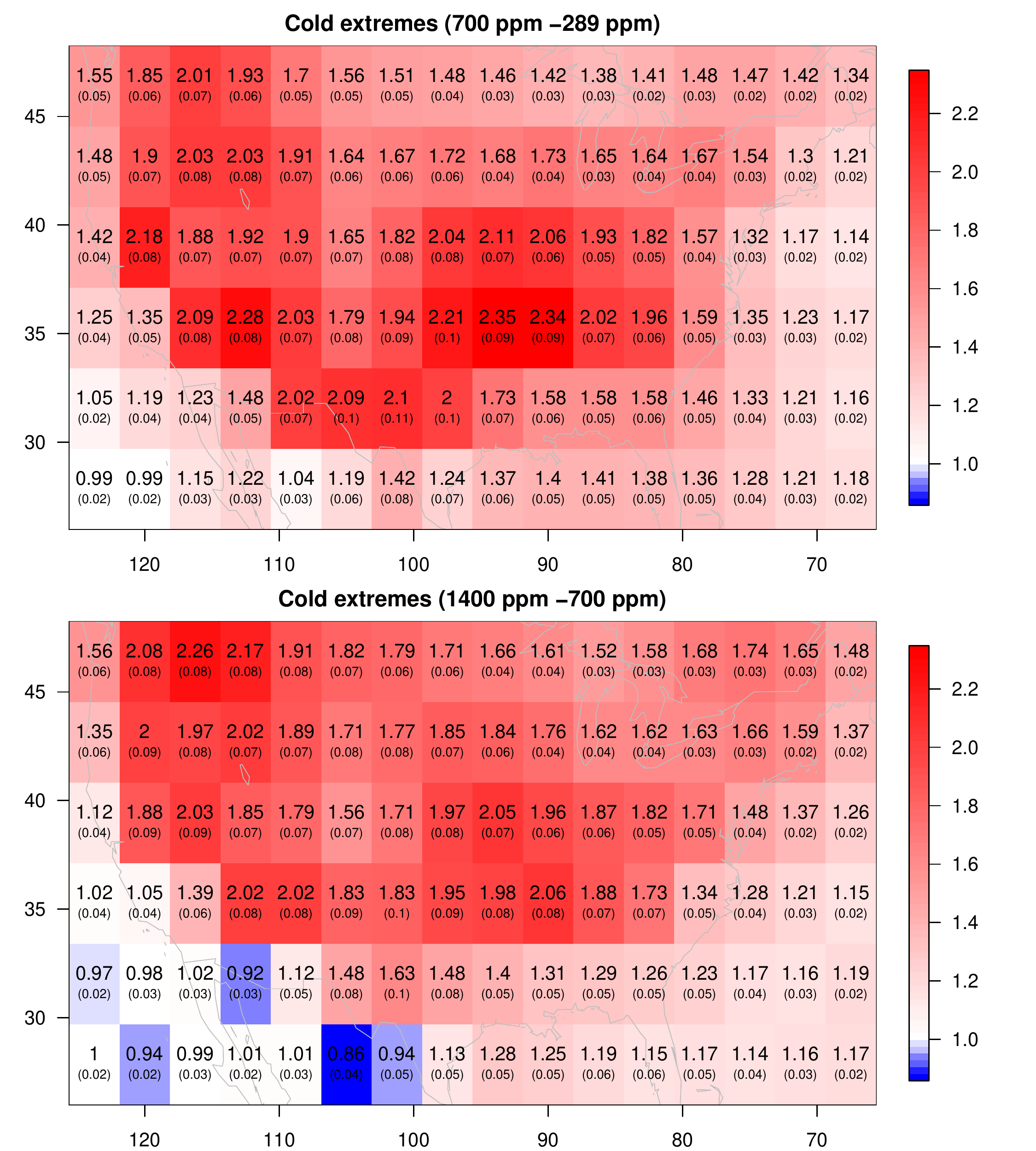}
\caption{As in Fig. \ref{fig:ratio_gev_mu_su} but for changes in the location parameter of cold extremes to changes in winter means. These ratios are substantially greater than 1 nearly everywhere in the contiguous United States.}
\label{fig:ratio_gev_mu_wi}
\end{figure}

Here we explore the relationship of changes in the location parameter of extremes to changes in the corresponding seasonal means (Fig.~\ref{fig:ratio_gev_mu_su} and Fig.~\ref{fig:ratio_gev_mu_wi}). We find that the changes in the location parameter of warm extremes generally follow the mean changes during summer (see Fig.~\ref{fig:ratio_gev_mu_su}). Changes in the location parameter of cold extremes, however, are usually larger. For most inland locations, changes in cold extremes are amplified by more than 50\% relative to changes in the average winter T$_{\text{min}}$ (see Fig.~\ref{fig:ratio_gev_mu_wi}).

\section{Changes in the shape parameter from pre-industrial to 1400 ppm} \label{App:xi_289_1400}
Here we present changes in the estimated shape parameter of extremes in the transition from pre-industrial to 1400 ppm climate states.  

\begin{figure}[H]
\centering
\includegraphics[width=5in]{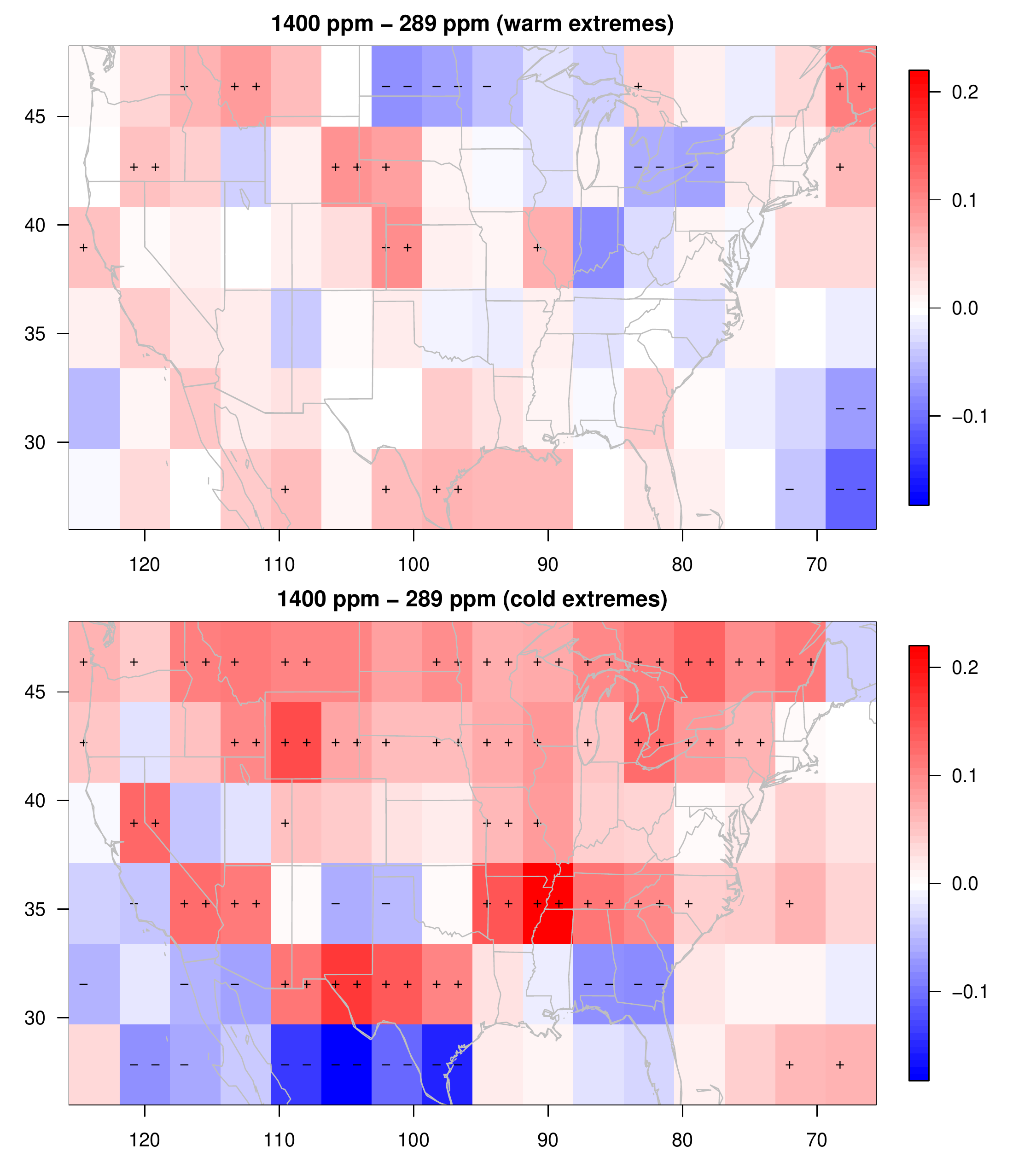}
\caption{Changes in the estimated CCSM3 GEV shape parameter in the transition from pre-industrial to 1400 ppm climate states. \textbf{Upper}: warm extremes. \textbf{Lower}: cold extremes. Changes in the shape parameter are in general significant for inland cold extremes.}
\label{fig:delta_xi_289_1400}
\end{figure}

\section{Comparsion of maximum likelihood (ML) and probability weighted moments (PWM) estimation} \label{App:PWM_ML}
\begin{figure}[H]
\centering
\includegraphics[width=5.75in]{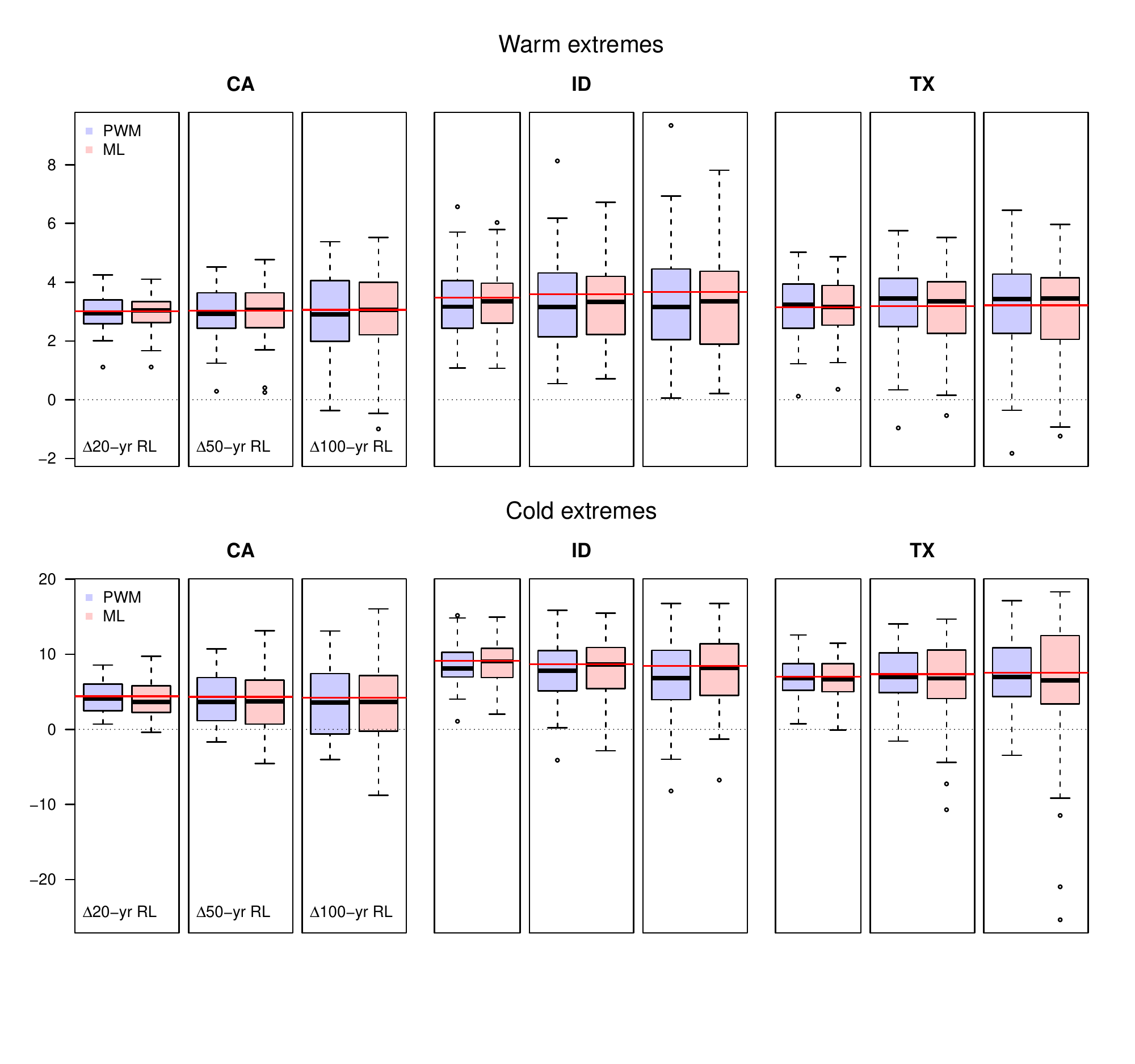}
\caption{Boxplots of estimated changes in 20, 50, and 100 year return levels using PWM (blue) and ML (red).}
\label{fig:pwm}
\end{figure}

Here we present the comparison of the estimates by fitting 20-year segments using ML and PWM, respectively. In general, the ML and PWM give very similar results (see Fig~\ref{fig:pwm}) except that the PWM avoid the unreliable estimates for cold extremes in TX location. The reader is referred to \cite{hosking1985} for the details of PWM.    
    
\end{appendices}

  \bibliography{ascmo.bib}

\end{document}